\documentclass[preprint,aip,cha]{revtex4-1}

\usepackage[english]{babel}
\usepackage{lineno}
\usepackage{graphicx}
\usepackage{dcolumn}
\usepackage{bm}
\usepackage{amsmath}
\usepackage{amssymb}
\usepackage{physics}
\usepackage{datetime}
\usepackage{hyperref}
\usepackage{bigints}
\usepackage{siunitx}
\usepackage{subcaption}
\usepackage{caption}
 \usepackage{todonotes}
\usepackage{color, colortbl}
\usepackage{cleveref}
\usepackage{multirow}
\usepackage{diagbox}
\usepackage{rotating}
\usepackage[normalem]{ulem}
\definecolor{Gray}{gray}{0.9}
\begin{document}

\title{ Drift-kinetic PIC simulations of plasma flow and energy transport  in the magnetic mirror configuration}
\author{\firstname{M.}~\surname{Tyushev}}%
\email{mikhail.tyushev@usask.ca}
\thanks{Corresponding author}
\author{\firstname{A.}~\surname{Smolyakov}}
\author{\firstname{A.}~\surname{Sabo}}
\affiliation{Department of Physics and Engineering Physics, University of Saskatchewan, Saskatoon SK S7N 5E2, Canada}
\author{\firstname{R.}~\surname{Groenewald}}
\author{\firstname{A.}~\surname{Necas}}
\author{\firstname{P.}~\surname{Yushmanov}}
\affiliation{TAE Technologies, Inc., 19631 Pauling, Foothill Ranch, CA 92610}

\begin{abstract}

Plasma flow and acceleration in a magnetic mirror configuration are studied using a drift-kinetic particles-in-cell model in the paraxial approximation, with an emphasis on finite temperature effects and energy transport. Energy conversion between electrons and ions, overall energy balance, and axial energy losses are investigated. The simulations of plasma flow, acceleration, and energy transport in the magnetic mirror are extended into the high-density regimes with implicit particle-in-cell simulations. It is shown that profiles of the anisotropic ion temperatures and heat fluxes obtained with the full drift-kinetic model compare favorably with the results of a fluid model, which includes collisionless ion heat fluxes beyond the two-pressure adiabatic equations. The effects of collisions on trapped electrons and the resulting impacts on electron temperature and electric field profiles are investigated using a model collision operator.  
\end{abstract}

\maketitle


\section{Introduction}

Plasma flow in magnetic mirrors with converging-diverging magnetic fields is of interest for fusion, plasma processing, and electric propulsion applications. The magnetic mirror is one of the oldest schemes for magnetic confinement, aiming to create conditions for controlled fusion. The geometry of a highly diverging magnetic field is the basis for advanced divertor configurations designed to handle the heat flow to the walls of the fusion reactor \cite{RyutovFST2005}. Related physical phenomena occur in the magnetic nozzle\cite{comment} and magnetic mirror (converging-diverging) configurations, which are used to convert internal plasma energy into the directed kinetic energy of the supersonic plasma to create thrust in plasma propulsion devices. All these applications require a good understanding of the mechanisms of plasma flow acceleration and energy transport in the magnetic mirror configurations of various settings, e.g., see Refs. \onlinecite{Morozov_V8,IvanovUFN2017} and references therein. A recent overview of the work in the context of electric propulsion can be found in Ref. \onlinecite{KaganovichPoP2020}.

With simple assumptions of quasineutrality and isothermal electrons, plasma flow in the converging-diverging magnetic mirror configurations is controlled by the robust profile of the electrostatic potential, which is uniquely determined by the magnetic field and the sonic condition at the mirror throat \cite{SmolyakovPoP2021,SaboPoP2022,JimenezPoP2022}.   More generally, accelerating potential structures may be generated due to the effects of the magnetic field profile and mechanical apertures, e.g. at the interface of the plasma source with the expansion region, \cite{CharlesAPL2003,CharlesAPL2007,SefkowPoP2009,SunPRL2005} as well as plasma pressure anisotropies in real and phase space,  e.g. due to presence of several species with different temperatures. These accelerating structures can be quasineutral or involve space-charge effects   \cite{AhedoPRL2009}. 

In weakly collisional plasmas typical for many applications, plasma expansion in the mirror magnetic field will naturally lead to the development of pressure anisotropy even if the plasma source is initially isotropic. The pressure anisotropies are strongly affected by reflections due to the magnetic and electric fields. Overall, in weakly collisional regimes, electron and ion dynamics in the magnetic mirror are non-local \cite{BoswellFP2015} due to the global nature of the electric field formed by global ambipolarity and quasineutrality constraints. Such a non-local electric field may also be coupled to the boundary conditions such as sheaths at the boundaries of finite-length systems. All these nonlinear and kinetic phenomena typically require numerical simulations.

Previously, the effects of a finite ion temperature on plasma acceleration in the magnetic mirror were studied using a fluid two-pressure adiabatic model and Boltzmann electrons \cite{SaboPoP2022}. Here, a full drift-kinetic model for ions and electrons is used to study the formation of the global electric field profile, energy transport, and resulting plasma acceleration.

The drift-kinetic model in the paraxial approximation used in our work is similar to the models used in Refs. \onlinecite{MartinezPoP2011,SanchezPSST2018,ZhouPSST2021,ChaconJCP2024}. In Ref. \onlinecite{MartinezSPoP2015}, the stationary distribution functions for electrons and ions as functions of the integrals of motion were used to find the electrostatic potential by numerical iterations, imposing the quasineutrality and ambipolarity (current-free) conditions and assuming empirically that the region of doubly trapped electrons in the expanding part is fully populated. With the above assumptions, the electron and ion temperatures and heat fluxes were calculated as moments of the distribution function. A similar model was used to analyze parametric dependences, in particular, the effect of heat fluxes in the ion and electron energy balance \cite{AhedoPSST2020}. The time-dependent Vlasov-Poisson model was used to study the effects of non-stationary electron trapping \cite{SanchezPSST2018}. The effects of collisions were studied in the time-dependent Boltzmann-Poisson model \cite{ZhouPSST2021} that included the Bhatnagar-Gross-Krook collision operator. The drift-kinetic paraxial nozzle was studied with an implicit, conservative particle-in-cell algorithm in Ref. \onlinecite{ChaconJCP2024}, employing novel boundary conditions to properly describe the expansion-to-infinity situation typical for propulsion applications.

Similar to previous works, in this paper, we address the formation of the electron and ion pressure anisotropy, electron, and ion energy fluxes, and their role in plasma flow acceleration with an emphasis on the energy balance and energy conversion between electrons and ions. The emphasis of our study is on the collisionless case,  and, therefore, in the base case, the region of trapped electrons in the expander is disconnected from the source region. In this collisionless case, the trapped region can be filled only by transient (non-stationary) processes and as a result of the numerical noise inherently present in PIC simulations. As a result, we observe a small number of electrons in the trapped region. In this limit,  we compare the results of our kinetic studies with the extended fluid model for ions \cite{Sabo2} that includes the collisionless heat fluxes beyond the two-pressure adiabatic model \cite{CGL1956}.
Whenever possible, we compare our results with previous analytical results \cite{SmolyakovPoP2021,SaboPoP2022} and numerical simulations\cite{MartinezPoP2011,SanchezPSST2018,ZhouPSST2021,MartinezSPoP2015} using Vlasov-Poisson and Boltzmann-Poisson models.

Our study is focused on open mirror fusion applications \cite{GotaNF2021,OnofriPoP2017,FrancisquezPoP2023,WethertonPoP2021,SoldatkinaNF2020} where the energy flux from the mirror is expected to be absorbed on the wall. Therefore, we take into account the absorbing wall and investigate how the mirror effects (mirror forces and particle reflections) affect the potential profile (including the sheath at the absorbing wall) and energy transport for large magnetic field expansion proposed to reduce the overall energy losses \cite{KonkashbaevJETP1978,RyutovFST2005}. 
Previous studies \cite{Pastukhov_1974,KonkashbaevJETP1978,RyutovFST2005}
emphasized the role of collisions (inside the plasma source) in forming the electron distribution function in the loss cone region. In our model, we consider a Maxwellian plasma source with a completely full loss cone -- the worst case from the perspective of energy losses. Collisions also affect particles in the expander region, in particular by providing a mechanism for electron trapping in that region. To evaluate these effects, we perform a separate study using the model collision operator for electron-neutral collisions.    

Our simulation model is described in Section II. Section III presents the results of the simulations demonstrating the effect of the ion acceleration, formation of anisotropic distribution functions for electrons and ions, finite ion temperature effects, energy transport, and energy losses. Section IV describes the comparison of the results from WarpX \cite{WarpX} and EDIPIC \cite{SydorenkoTh2006} and the results of implicit simulations with EDIPIC. The effects of collisions are considered in Section  V. Section VI presents the summary and discussion. Appendix A presents the Poisson equation in the paraxial model. Boundary conditions are discussed in Appendix B. Appendix C summarizes the extended hydrodynamic equations for ions.    
\section{Simulation model}\label{sec_model}

\subsection{Drift kinetic equation and paraxial approximation in PIC}

\label{sec_model}

The motion of ions and electrons in our model is considered in the drift-kinetic approximation, assuming that the characteristic frequencies of all relevant processes are much smaller than the ion/electron cyclotron frequencies,  $\omega \ll \omega _{ci} \ll \omega _{ce}$, and particles' Larmor radii are much smaller than the transverse length scale, so the motion of the guiding centers represents the particle dynamics.

Full equations of motion in the drift-kinetic approximation have the form
\cite{Sivukhin1965}:
\begin{equation}
\frac{d\mathbf{r}}{dt}=v_{\parallel }\mathbf{b}+\mathbf{V}_{E}+\mathbf{V}%
_{d},  \label{dRdt}
\end{equation}%
\begin{equation}
\frac{dv_{\parallel }}{dt}=\frac{q}{m}\mathbf{E}\cdot \mathbf{b}+\frac{1}{2}%
v_{\perp }^{2}\nabla \cdot \mathbf{b}+v_{\parallel }\mathbf{V}_{E}\cdot
\nabla \ln B,
\end{equation}
\begin{equation}
\frac{dv_{\perp }^{2}}{dt}=-v_{\perp }^{2}v_{\parallel }\nabla \cdot \mathbf{%
b}+v_{\perp }\mathbf{V}_{E}\cdot \nabla \ln B,  \label{dvperpdt}
\end{equation}%
where $\mathbf{b}=\mathbf{B}/B$ is a unit vector along the magnetic field, $q$ is the particle charge, $m$ is the particle mass, $\mathbf{V}%
_{E}=\mathbf{E}\times \mathbf{B}/B^{2}$ is the $\mathbf{E}\times \mathbf{B}$
drift, and
\begin{equation}
\mathbf{V}_{d}=\frac{v_{\perp }^{2}/2+v_{\parallel }^{2}}{\omega _{c}}%
\mathbf{b}\times \nabla \ln B,
\end{equation}
is the curvature and magnetic gradient drift, which are equivalent  to the diamagnetic drift due to the pressure gradients.    Here $v_{\parallel }$ and $v_{\perp }$ are particles' velocities along and perpendicular to the magnetic field. In the axisymmetric geometry of the magnetic nozzle, the $\mathbf{E} \times \mathbf{B}$ and diamagnetic drifts are azimuthal (in the symmetry direction) and can be dropped in the electrostatic approximation. The collisionless drift-kinetic equation for the distribution function of guiding centers $f=f(\mathbf{r},t,v_{\parallel },v_{\perp })$, where $\mathbf{r}$ is a radius vector of a guiding center position, is written in the form
\begin{equation}
 \frac{\partial f}{\partial t}+\nabla \cdot \left( \frac{d\mathbf{r}%
}{dt}f\right) +\frac{\partial }{\partial v_{\parallel }}\left( \frac{%
dv_{\parallel }}{dt}f\right) +\frac{\partial }{\partial v_{\perp }^{2}}%
\left( \frac{dv_{\perp }^{2}}{dt}f\right) =0.
\label{dke}
\end{equation}%
Note that the drift equations (\ref{dRdt}-\ref{dvperpdt}) conserve the phase space volume in the form%
\begin{equation}
\ \nabla \cdot \left( \frac{d\mathbf{r}}{dt}\right) +\frac{\partial }{%
\partial v_{\parallel }}\left( \frac{dv_{\parallel }}{dt}\right) +\frac{%
\partial }{\partial v_{\perp }^{2}}\left( \frac{dv_{\perp }^{2}}{dt}\right)
=0.
\end{equation}
Equations (\ref{dRdt}-\ref{dvperpdt}) have two conserved integrals, energy and magnetic moment (adiabatic invariant) $\mu = mv_{\perp }^{2}/2B$, $d\mu /dt = 0$. It is convenient to write the drift-kinetic equation in $(v_{\Vert }$, $\mu )$ variables for $f=f\left( \mathbf{r},t,v_{\parallel },\mu \right)$
\begin{equation}
 \frac{\partial f}{\partial t}+v_{\parallel }\mathbf{b}\cdot \nabla
f+\frac{dv_{\parallel }}{dt}\frac{\partial f}{\partial v_{\parallel }}=0,
\label{dke2}
\end{equation}%
where%
\begin{equation}
\frac{dv_{\parallel }}{dt}=\frac{q}{m}\mathbf{E}\cdot \mathbf{b-} \frac{\mu}{m} \mathbf{%
b\cdot }\nabla B.  \label{dvpar}
\end{equation}%
Thus, the motion of the guiding center becomes one-dimensional along the magnetic field line. The particle density is found from $f=f\left( \mathbf{r},t,v_{\parallel },\mu \right)$ as
\begin{equation}
n=\frac{2\pi B}{m}\int f\left( \mathbf{r},t,v_{\parallel },\mu \right)
dv_{\Vert }d\mu .  \label{nb}
\end{equation}
In the paraxial approximation, the electric field is assumed to be along the magnetic field line $\mathbf{E} = E_{\parallel }\mathbf{b}$, $E_{\parallel } = \mathbf{E} \cdot \mathbf{b}$, so that the Poisson equation can be written in the form
\begin{equation}
\nabla \cdot \mathbf{E}=\ \nabla _{\parallel }E_{\parallel }+E_{\parallel
}\nabla \cdot \mathbf{b}=\nabla _{\parallel }E_{\parallel }-E_{\parallel
}B^{-1}\nabla _{\parallel }B=4\pi e\left(
n_{i}-n_{e}\right) ,  \label{parp}
\end{equation}%
where $\nabla _{\parallel }=\mathbf{b}\cdot \nabla$. Alternatively, equation (\ref{parp}) can be obtained by averaging the two-dimensional Poisson equation over the flux tube cross-section as shown in Appendix A.

Therefore, in the paraxial approximation, we can model the fully kinetic self-consistent plasma dynamics in the magnetic nozzle with a one-dimensional code by including the electric field and mirror forces according to Eq. (\ref{dvpar}), $\nabla_{\parallel}\simeq \partial /\partial z$, $E_{\parallel}\simeq -\partial \phi /\partial z$, and solving the modified Poisson equation (\ref{parp}). This model is equivalent to the model used in Ref. \onlinecite{ZhouPSST2021} where it was solved by using direct integration of Eq. (\ref{dke2}). Here, similarly to Ref. \onlinecite{ChaconJCP2024}, we use the Particle-in-Cell approach to solve equations (\ref{dke2}) and (\ref{parp}). The particles' perpendicular velocity is adjusted following the particle position and the magnetic moment conservation. In the one-dimensional model, the variation of the magnetic field is included by the Jacobian of the transformation to the $({\mathbf r}, v_\Vert, \mu)$ variables \cite{EBERSOHN2017,ChaconJCP2024}. This gives the linear relation of plasma density to the magnetic field when the distribution function is written in $(v_\Vert, \mu)$ variables, equation (\ref{nb}). This dependence takes into account plasma compression (or expansion) when particles follow the converging (diverging) magnetic field, so that particle flux is conserved for each species ($\alpha$), $n_\alpha V_{\alpha \parallel}/B = const$, where $V_{\alpha \parallel}$ is a flow velocity along the field lines. In the PIC setup, this is achieved by the variation of the cell volume \cite{EBERSOHN2017} giving for the density (electron and ion)
\begin{equation}
n_{j}^{^{\prime }}=n_{j}B_{0}/B_{j} \equiv n_j / \Hat{B}_j,
\label{Eq:Adjust}
\end{equation}
where $\Hat{B}_j = B_j / B_0$ or $\Hat{B}(z) = B(z) / B_0$ (for the continuous representation) is the magnetic field normalized to the magnetic field at $z=0$.


\subsection{Plasma source and Maxwellian reflux injection model}

\begin{figure}[htp]
\centering
\captionsetup{justification=raggedright,singlelinecheck=false}
\includegraphics[width=1\textwidth]{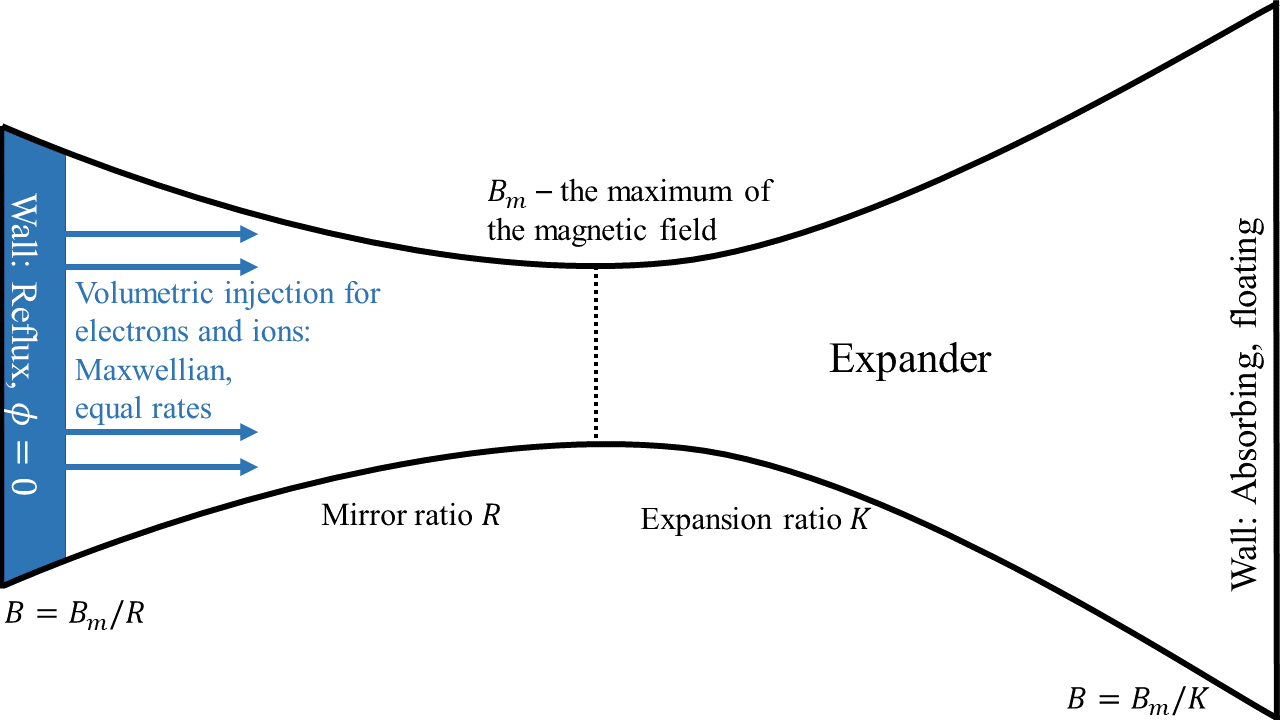}
\caption{A  schematic diagram of the setup of simulations.}
\label{fig:Diagram}
\end{figure}

A general setup of the simulations is represented in Fig.\ref{fig:Diagram}.  In our model, we do not include ionization as the actual source of plasma. 
Our left boundary represents a junction of the simulated magnetic nozzle with the plasma source where plasma is generated and heated. It is assumed that the residence time of particles in the plasma source is large compared to collision times, or equivalently, the mean free collision path is shorter than the length of the source.  The electrons and ions are injected from the left boundary with the Maxwellian distribution and given temperatures assumed in the source. The electrons and ions are injected at equal rates. 
Any particles that are reflected back to the boundary (and crossing it from the right to the left) are reflected again back into the nozzle but resampled from the same Maxwellian distribution function assumed in the source for each species. Thus, the number of injected particles (per unit of time) in each species remains the same. In this setup, we may inject energy but not the charge, so that the plasma injection is current-free {(ambipolar)}. It {remains} current-free along the nozzle due to the flux conservation condition $n_\alpha V_{\alpha \Vert}/B=const  $  inherent in our model for each species.

The electrons and ions are injected volumetrically in the narrow region near the left wall.  Particle positions are sampled from the distribution $z/L = 0.1 \times U^{1/4}$, where $U$ is a uniform number from $[0;1)$ and $L$ is the system length (1.2 m)


As proposed in Ref. \onlinecite{CartwrightJCP2000} and used in many PIC simulations,  electrons and ions are injected from the flux Maxwellian distribution in the $z$ direction perpendicular to  the wall
\begin{equation}
       f_{z} =  \frac{2 \Gamma v_z}{v_{T_{e/i}}^2} \exp(-\frac{v_z^2}{v_{T_{e/i}}^{2}}), \hspace{0.5 cm} v_z>0,
\label{FluxMaxwell}
\end{equation}
where   $\Gamma $  is the injected particle flux. In the directions parallel to the wall, we sample particles from a 1D Maxwellian distribution, separately for each direction,
\begin{equation}
f_{x/y} = \frac{1}{\sqrt{\pi} v_{T_{e/i}}} \exp(-\frac{v^2_{x/y}}{v_{T_{e/i}}^{2}}), 
\label{Maxwell}
\end{equation}
  $v_{T_{e/i}} = \sqrt{2T_{e/i}/m_{e/i}}$.  

 We do not employ any injection control nor impose any quasineutrality condition as it is done in alternative approaches in Refs.   \onlinecite{ZhouPSST2021,ChaconJCP2024}: the rates of injection for ions and electrons are fixed and kept equal so that the plasma flow is ambipolar (current-free). In this model, the only fixed parameters are the injection rate (equal for ions and electrons) and temperatures of the Maxwellian distributions from which injected particles are sampled. The resulting densities of electrons and ions throughout the nozzle (including the region near the left boundary) and actual energy distribution (the ion and electron effective temperatures) are established as a result of self-consistent particle and electric field dynamics including particles reflections from the magnetic mirror and the electric field. As it was noted above, a fraction of injected particles is reflected back to the wall by the magnetic field, and the electric field formed self-consistently. Such particles are then reinjected with their velocities sampled from the original Maxwellian distribution of the same temperature. In what follows, this setup is referred to as the Maxwellian reflux model \cite{SchwagerPFB1990}.   

This simulation and injection model setup aims to represent plasma flow into the nozzle from the long plasma source with a given source of particles which is fixed by the chosen injection rate. The energy flux for each species is not fixed in this model: the particles gain energy during reflections back into the source and subsequent resampling.  The energy flux into the nozzle per one ion-electron pair  is a figure of merit of the mirror confinement systems and is one of the parameters of interest in our work.




\subsection{The drift-kinetic model in EDIPIC and WarpX and boundary conditions}

In our work, we have used two open-access Particle-In-Cell (PIC) codes: the one-dimensional Electrostatic Direct Implicit Particle-In-Cell (EDIPIC) code developed by D. Sydorenko \cite{SydorenkoTh2006}, available at \url{https://github.com/PrincetonUniversity/EDIPIC}, and WarpX (Version 23.09), a massively parallel PIC code for kinetic plasma simulations \cite{WarpX}, \url{https://github.com/ECP-WarpX/WarpX}. We have modified both codes into the drift-kinetic versions. The EDIPIC is a momentum-conserving code, and for WarpX we used the momentum-conserving option (it also has an energy-conserving option). The results from the two codes provide a useful benchmark comparison. Another important goal was to compare the results of the implicit EDIPIC code with the explicit WarpX, as elaborated in Section IV. The modified EDIPIC and WarpX codes can be found here: \url{https://github.com/Maknagens/EDIPIC-Drift-Kinetic} and \url{https://github.com/Maknagens/WarpX-1D-Drift-kinetic}

The magnetic mirror force $-\mu \nabla B$ was introduced prior to the standard electric push in both codes. Concentrations of electrons and ions were adjusted according to the cell position on the magnetic field line, Eq. (\ref{Eq:Adjust}). The Poisson equation was modified with the additional term, as shown in Eq. (\ref{parp}). By approximating this equation using central differences, we derived the following form
\begin{equation}
    \phi^{j-1} - 2\phi^{j} + \phi^{j + 1} + (B^{-1}\nabla_{\parallel} B)^j \Delta_{\parallel} z (\phi^{j-1} - \phi^{j+1}) / 2 = 4\pi e (n_i - n_e)^j \Delta_{\parallel} z^2.
    \label{numphi}
\end{equation}
In this form, it is a tri-diagonal matrix that can be efficiently handled using the Thomas algorithm.  The magnetic field inhomogeneity effects in Eq. (\ref{numphi}) were added to the Thomas algorithm used in the EDIPIC code. For WarpX we directly solve Eq. (\ref{numphi}) with the Thomas algorithm instead of using a prebuilt multigrid solver. This allows us to solve the Poisson equation exactly and implement the magnetic field effect. Additionally, the Maxwellian injection and reflux algorithms for the left wall were implemented in both codes. Finally, since WarpX lacked a default floating wall boundary condition, we included it analogously to the implementation in EDIPIC. All modifications to WarpX were made using the Python Particle-In-Cell Modeling Interface (PICMI), distributed with WarpX.  We use the floating wall at the right wall, $z=L$, and $\phi=0$ at the left wall, $z=0$.


\subsection{Magnetic field profile   and other simulation parameters}



The overall setup of the simulations is shown in \ref{fig:Diagram}. 
We use the following model for the magnetic field with different mirror and expansion ratios
\begin{equation}
\begin{cases}
B(z') = (1 + 4(R-1)  (z' - 0.5)^2)^{-1}, z' \leq 0.5 \\
B(z') = (1 + 4(K-1)  (0.5 - z')^2)^{-1}, z'\geq 0.5
\end{cases},
\label{Eq:MagField}
\end{equation}
where $R$ and $K$ are the mirror and expansion ratios correspondingly, $z'=z/L$ is the axial position normalized to the system length $L$, $z'=0.5$ is the location of the mirror throat. Note that while the magnetic field and the first derivative in the converging and diverging part are matched across $z'=0.5$, there is a discontinuity in the second derivative. The consequences of this discontinuity are discussed below in Section III.A.

As was mentioned in the previous subsection, the left wall has a fixed potential $\phi = 0$.  
  The right wall is absorbing and floating,  which together with a global current-free condition guarantees the achievement of a stationary state.  We have also tested alternative boundary conditions which are discussed in Appendix B.

 For the base case parameters, 
 we consider the injection temperatures $T^e_0 = 300 \: eV$ for electrons, and $T^i_0 = 30 \: eV$ for Hydrogen ions, with the injected flux (in units of the current) of $I_{inj} = 10$ A/m$^2$.   Some simulations are performed at different ion temperatures (30/300/600 eV), as it is specified in the next sections. 
 
 The base case magnetic field mirror and expansion ratios are $R=10$ and $K=50$. The profile of the field from Eq. (\ref{Eq:MagField}) is shown in  Fig. \ref{fig:ShwagMagConc}d.  The size of the simulation domain is $L=1.2$ m. The PPC number (Particle Per Cell) is 1000 on average. The Debye length is resolved in the whole domain -- in the region of the largest concentration (at the left wall), we have 1.9 cells per Debye. There are 1179 cells distributed evenly across the domain, and the time step is $2.5\times10^{-11}$ s. Free streaming condition is $v_{T^e_0} \Delta t / \Delta x = 0.25$, where $v_{T^e_0} = \sqrt{2 T^e_0 / m_e}$. For the base case, we inject 7.6 particles (on average) at every time step. One particle is injected randomly when the generated random number is less than 0.6. There are no particles in the simulation domain at the start of the simulations.

 The simulations reach a steady state at $\sim$ 100 microseconds for $R=10$ and $K=50$ regime. The steady state is reached after many ions bounce between the left wall and reflections in the Yushmanov potential, so the injected ion/electron fluxes equilibrate at the right wall. The steady-state energy balance is reached approximately at the same time. The number of reflections required to reach steady-state increases with $R$ since the injection rate is the same and the density in the source region increases due to improved confinement.
 
 The profiles of the potential, ion velocity, ion concentration, etc., shown in what follows are averaged with the $3$ microseconds time frame with $100$ snapshots to suppress noise. 

 \subsection{Quasineutral hybrid model}

 Our main results in this paper are based on full kinetic descriptions for ions and electrons. It is of interest to compare these with the quasineutral approach, in which the condition $n_i = n_e$ is used together with the Boltzmann relation $n = n_0 \exp(e\phi/T_e)$ for the electron density, where $n_0$ represents the ion concentration at the left wall, and the electron  temperature $T_e$  is assumed constant. Using the ion density obtained from our kinetic solution and assuming isothermal electrons, one can extract the potential as $e\phi = T_e \ln(n/n_0)$ and subsequently move the ions according to the electric field from this expression. This approach shall be referred to below as the hybrid model and was used to characterize basic features of plasma acceleration in the magnetic nozzle in Ref. \onlinecite{JimenezPoP2022}.



\section{Results}

\subsection{The base case analysis: plasma acceleration in the magnetic mirror and the plasma flow in the uniform magnetic field}

\begin{figure}[htp]
\centering
\captionsetup{justification=raggedright,singlelinecheck=false}
\includegraphics[width=0.8\textwidth]{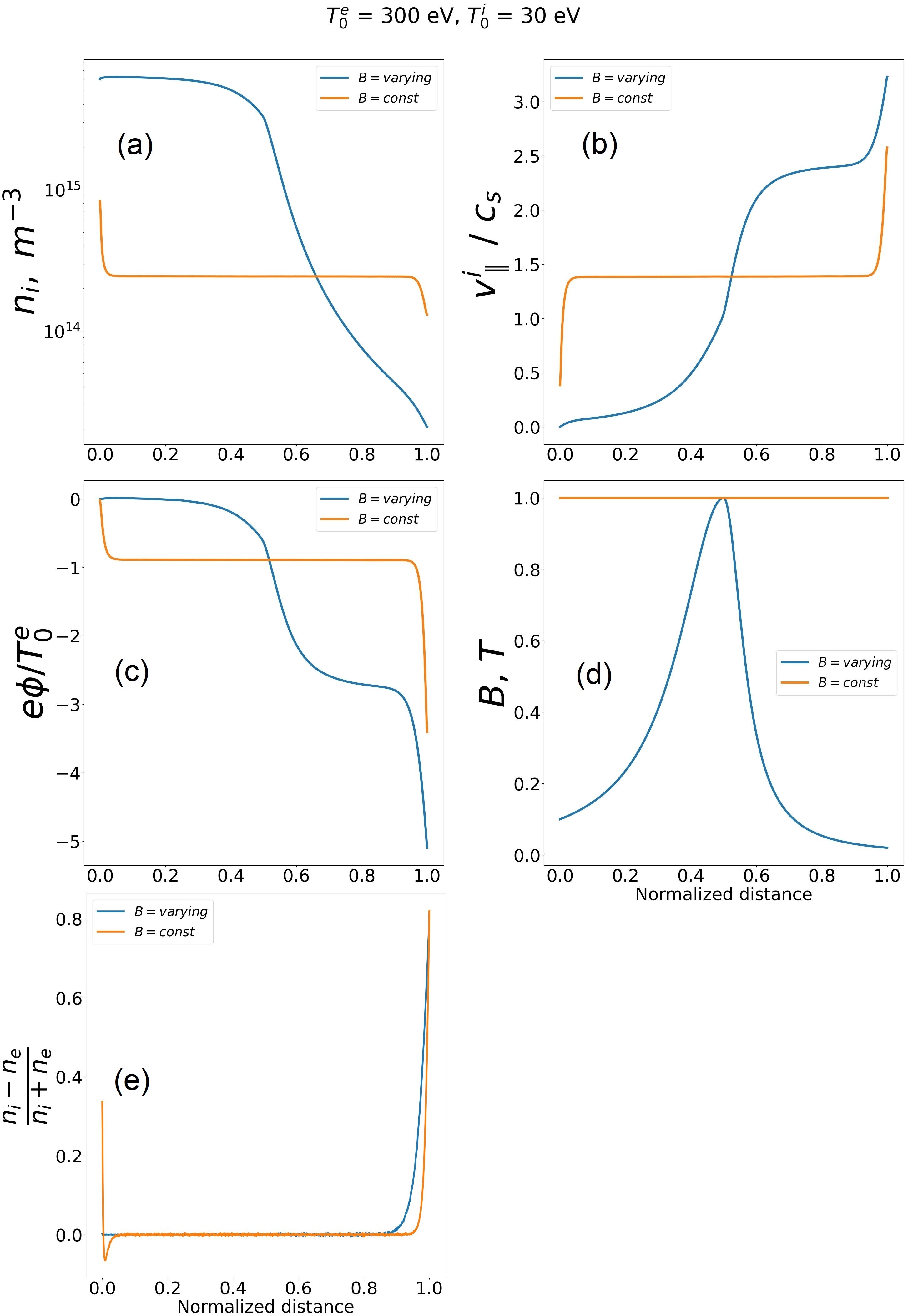}
\caption{Plasma density (a), ion velocity (b), potential profile (c), and deviation from quasineutrality (e) for plasma acceleration in the mirror magnetic field (d) with $R=10$, $K=50$. Here $c_s = \sqrt{T^e_0/m_i}$, $T^e_0$, and $T^i_0$ are the injection values of the electron and ion temperatures. Plasma parameter profiles for the uniform magnetic field are shown in orange providing a benchmark comparison with results in Ref.  \onlinecite{SchwagerPFB1990}. Note that the source sheath existing in the uniform field, practically disappears in the mirror field, Fig. (e).}
\label{fig:ShwagMagConc}
\end{figure}

To highlight the effects of the mirror field on plasma acceleration we first consider the plasma flow in the uniform magnetic field for the same injection model, boundary conditions, and $T_0^e=300$ eV and $T_0^i=30$ eV. This comparison,  shown in  Fig.  \ref{fig:ShwagMagConc}, also provides a useful check of our results with respect to the previous work in Ref.  \onlinecite{SchwagerPFB1990}, where the plasma flow under a similar injection model was considered in one-dimensional problem without magnetic field. In our model, the latter corresponds to the case of the uniform magnetic field.
A notable feature of plasma flow in the uniform magnetic field is the formation of the source sheath at the left wall. The source sheath occurs as a result of the higher electron thermal velocity (respectively the higher particle flux)  when the electrons and ions are injected at equal rates. The positive electric field is then generated near the injection region retarding the electrons. As a result,  the ions are accelerated supersonically in the narrow (of the order of a few Debye length) transition layer of the source sheath between the plasma source and quasineutral plasma region as it is visible on the density, ion velocity, and potential profiles in  Figs. \ref{fig:ShwagMagConc}a,b, and c.  The potential drop in the source sheath depends on the ratio of the ion to electron injection temperatures and will decrease for larger ion temperatures \cite{SchwagerPFB1990} and even may become positive changing the electric field to negative, e.g. as it was shown in Ref. \onlinecite{JimenezPoP2022}.  The rapid and discontinuous transition at the boundary of the plasma source and expansion region was also detected in experiments, e.g. Ref. \onlinecite{CharlesAPL2003}.  


Plasma streaming toward the absorbing material wall has been considered in numerous conditions for fusion and low-temperature plasma applications, including the classical sheath problem at the plasma-wall interface. In the simplest model, a plasma boundary sheath is formed to equilibrate ion and electron fluxes to the floating wall, leading to the Bohm condition $V_i>c_s$ at the sheath entrance. In the standard model, the ions are assumed to become supersonic due to the acceleration in the quasineutral pre-sheath region. In the case of a uniform magnetic field, similar to Ref.\onlinecite{SchwagerPFB1990},   the collector sheath is formed near the right (absorbing) wall, even though the ions are accelerated supersonically in the source sheath, as seen in Figs. \ref{fig:ShwagMagConc}b; the electric field is zero across the quasineutral region, and there is no pre-sheath, see Fig. \ref{fig:ShwagMagConc}c. 

The source sheath practically disappears for the converging-diverging mirror configuration, but the collector sheath remains,   Fig. \ref{fig:ShwagMagConc}c and e.   
The case of the converging-diverging magnetic field demonstrates improved plasma confinement due to the mirror effect: one can observe a plasma density increase in the source region, Fig.  \ref{fig:ShwagMagConc}a. Also, one can note an increased value of the total potential drop across the mirror pointing to the larger number of reflected electrons and thus to the lower electron energy losses, Fig.  \ref{fig:ShwagMagConc}c.   Improved confinement is inherently coupled to plasma acceleration, which has a classical feature of the acceleration in the magnetic de Laval nozzle where plasma flow becomes supersonic at the sonic point located at the nozzle throat. The plasma flow velocity is monotonically increasing along the nozzle with the sonic point close to the location of the maximum of the magnetic field. This is the point where the plasma flow is equal to the local ion sound velocity, as shown in Fig. \ref{fig:ShwagMagConc}b, where Mach velocity $M\equiv V_\Vert^i/c_s$ and potential are normalized with the electron temperature of the injection, $T^e_0$, and $c_s=\sqrt{T^e_0/m_i}$.
The sonic point condition regularizes the singularity that may occur at this point in the quasineutral plasma. As it was shown in our previous work \cite{JimenezPoP2022}, our simulations here further demonstrate that the transonic accelerating velocity profile is a robust attractor for plasma flow in the magnetic mirror.

 It is worth noting that the exact expression for  the ion sound velocity depends on the used plasma model.  
It is easy to define the ion sound velocity for isothermal electrons and cold ions \cite{SmolyakovPoP2021} and for finite ion temperature with CGL model\cite{SaboPoP2022}. However, additional effects such as heat fluxes, ionization, and charge-exchange modify the definition of the ion-sound velocity and the location of the sonic point condition \cite{AhedoPoP2002,SaboPoP2022}.     
 
Nevertheless, one can see from  Fig. \ref{fig:ShwagMagConc}b that the sonic point condition $M=1$ at the maximum magnetic field with $\partial B / \partial z = 0 $ is well satisfied in our base case when the contribution of the ion temperature to the ion sound velocity can be neglected. One notes a small jump in the velocity gradient that occurs at the sonic point, Fig. \ref{fig:ShwagMagConc}b.   It was shown previously that the velocity gradient at the sonic point is determined by the second derivative of the magnetic field at the maximum, Eq. (14) in Ref. \onlinecite{SmolyakovPoP2021}. In our model, the second derivative of the magnetic field in Eq. (\ref{Eq:MagField}) is not continuous at the mirror throat. This explains the jump in the velocity in Fig. \ref{fig:ShwagMagConc}b at $z^\prime=0.5$, as predicted by the analytical theory\cite{SmolyakovPoP2021}.  

In the mirror field, there are two complementary effects leading to plasma acceleration. One is the apparent acceleration due to the reflection by the mirror force on the left side from the magnetic field maximum. This is simply the filtering effect: particles with low values of parallel (axial) velocity are reflected by the mirror, so the passing particles on average have larger net axial velocity. Such passing particles are further accelerated by the mirror force on the right side from the magnetic field maximum. Thus, apparent acceleration is simply a ballistic effect in the mirror field and does not involve the electric field. True acceleration of ions is related to the electric field, which occurs due to the plasma expansion and associated density gradient. The generated electric field accelerates ions and reflects electrons. In this process, the electron (thermal) energy is converted into the kinetic energy of accelerated ions. Note that for the base case parameters here, $T_0^e=300$ eV and $T_0^i=30$ eV, the ion temperature is much lower than that for the electrons, and the ion thermal effects are not significant.

\begin{figure}[htp]
\centering
\captionsetup{justification=raggedright,singlelinecheck=false}
\includegraphics[width=1\textwidth]{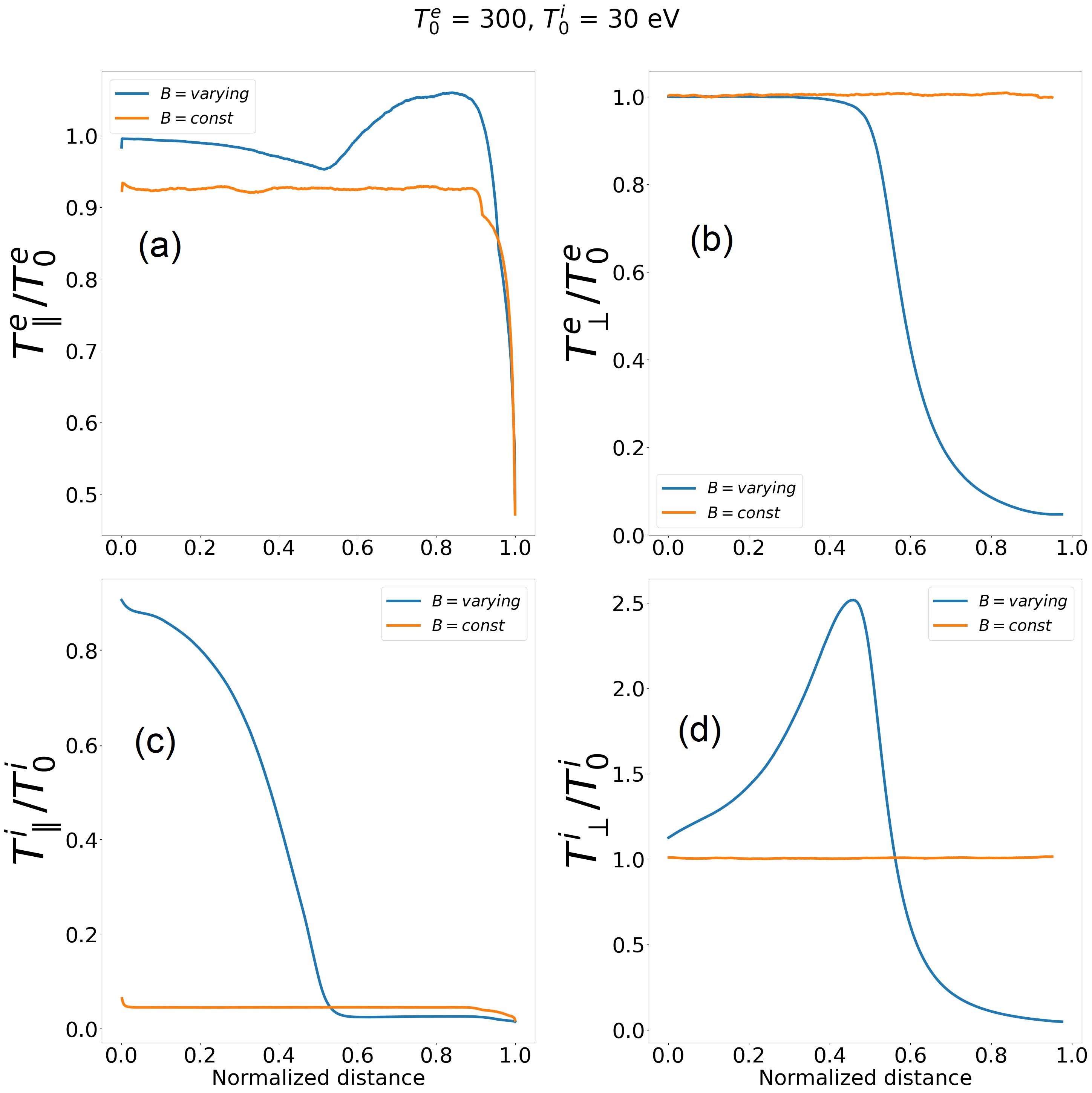}
\caption{Electron (a and b) and ion (c and d) and temperature profiles for the plasma flow in the mirror ($R=10$, $K=50$) and uniform magnetic field, $T^e_0$ and $T^i_0$ are the injection values of the electron and ion temperatures.}
\label{fig:ShwagMagTe}
\end{figure}

The conversion of the electron thermal energy into the ion kinetic energy (facilitated by the electric field) can be seen from the electron temperature profiles in Figure \ref{fig:ShwagMagTe}. The perpendicular electron temperature drops drastically in the expander region for $z^\prime >0.5$: the electron mirror force acceleration in this region occurs at the expense of the electron perpendicular energy.   Since the amount of the converted energy depends on the energy of passing electrons, and due to the asymmetry of the magnetic field (mirror region -- $R=10$ and expansion region -- $K=50$) it   leads to an increase of parallel temperature to $ \sim 1.1 T^e_0$ 



The behavior of the ion temperature is different from that of electrons: in Fig. \ref{fig:ShwagMagTe}d for the evolution of the ion perpendicular energy, one can see a \sout{weak} trend suggested by the conservation of the perpendicular adiabatic moment for individual particles. The increase of $T^i_\perp$  in the converging part of the mirror is much weaker than expected from the perpendicular CGL (Chew-Goldberger-Low) \cite{CGL1956} adiabatic constant, $S_\perp=p_\perp/nB$, $p_{\bot }=(m/2)\int v_{\bot }^{2} fd^{3}v$, so that $S_\perp$  is not conserved along the length of the mirror. The ion parallel temperature drops dramatically in the expander region, as seen in Fig. \ref{fig:ShwagMagTe}. One can associate this with the strong density and magnetic field dependence in the parallel CGL adiabatic constant, $S_\parallel=p_\parallel B^2/n^3$, $p_{\Vert }=m\int {v^{\prime }_\Vert}^2 fd^{3}v$, $v^{\prime }_\Vert$ is random velocity $v_\Vert =V_\Vert + v^{\prime }_\Vert$). In general, however, both two-pressure adiabatic invariants $S_\perp$ and $S_\parallel$ are not conserved due to the presence of the parallel heat fluxes. The role of the ion heat fluxes is discussed in further detail in Section III.E.

\subsection{ Anisotropy of the ion and electron distribution functions; trapped electrons}

\begin{figure}[htp]
\centering
\captionsetup{justification=raggedright,singlelinecheck=false}
\includegraphics[width=1\textwidth]{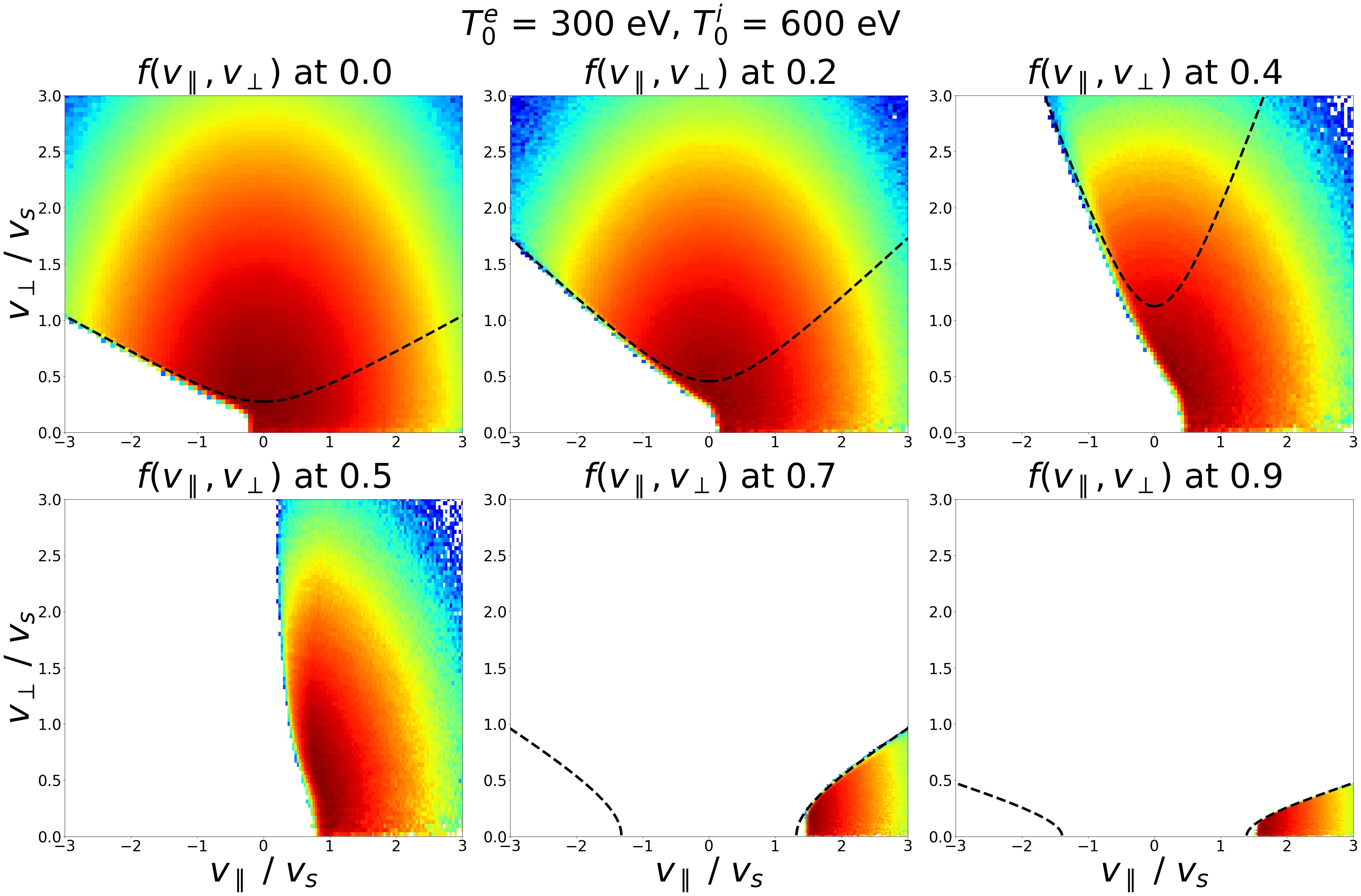}
\caption{The 2D map of the ion distribution function for several axial $z^\prime$ positions in the mirror with $R = 10, K=50$. The distribution function is averaged over the interval $[z^\prime,z^\prime+\Delta z^\prime]$, $\Delta z^\prime = 0.02$ for each location. The black dashed line shows the boundary of the ion loss cone modified by the electric field, Eq. (\ref{Eq:MagCond} represents the magnetic field ) confinement.  }
\label{fig:KineticR10_i}
\end{figure}

\begin{figure}[htp]
\centering
\captionsetup{justification=raggedright,singlelinecheck=false}
\includegraphics[width=1\textwidth]{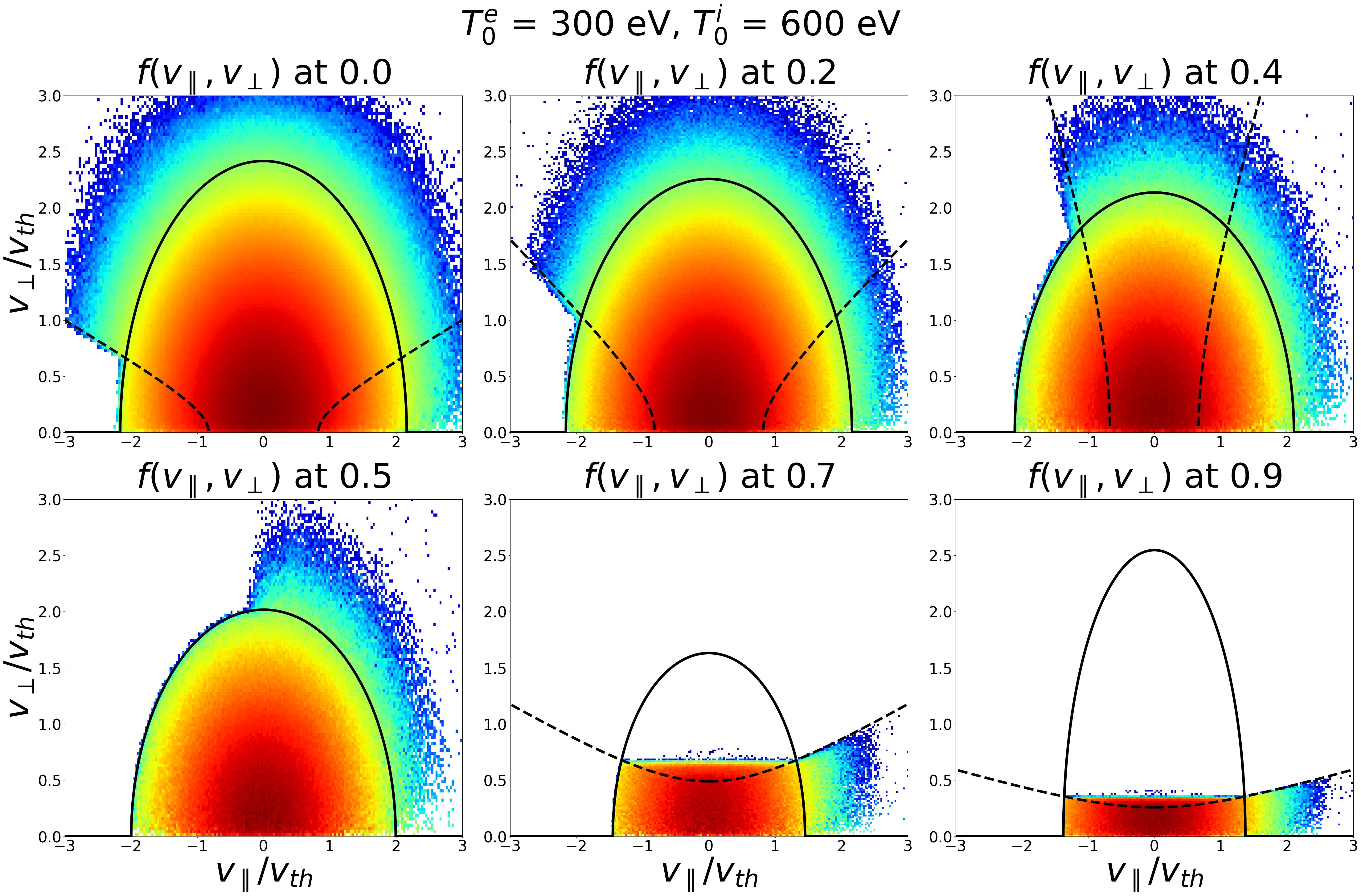}
\caption{The 2D map of the electron distribution function for several axial $z^\prime$ positions in the mirror with $R = 10,K=50$. The distribution function is averaged over the interval $[z^\prime,z^\prime+\Delta z]$, $\Delta z^\prime = 0.02$ for each location. The solid line shows the boundary of electrons  reflected by the electric field,  Eq. (\ref{Eq:PotCond}) and the dashed line -- electrons reflected by the magnetic field, Eq. (\ref{Eq:MagCond}).}
\label{fig:KineticR10_e}
\end{figure}

The development of anisotropy in electron and ion pressures is a natural result of collisionless plasma expansion in the magnetic field. Anisotropy of the ion and electron distribution can be well characterized by the two-dimensional maps in $v_\perp, v_\parallel$ space, as shown in Figs. \ref{fig:KineticR10_i}  
and \ref{fig:KineticR10_e} for different axial locations in the mirror with $R=10,K=50$.   The distribution function is averaged over the interval $[z^\prime,z^\prime+ \Delta z^\prime ]$, $\Delta z^\prime =0.02$ for each location.

The ion distribution function at $z^\prime=0$ is an isotropic Maxwellian distribution with an empty loss cone for $v_\Vert<0$. Except for the particles with low energy, in the vicinity of the origin $v_\Vert \simeq v_\perp \simeq 0$, the shape of the ion loss cone at $z^\prime=0$ is mostly determined by the reflections from the magnetic mirror with $R=10$. Further down along the axial direction, the ions gradually form the beam distribution due to the acceleration in the {electric field}. and the magnetic field effects due to the conversion of $v_\perp$ into $v_\parallel$ following the conservation of $\mu$.

The electron distribution function is more complex. The electric field reflects a fraction of the electrons, modifying the electron loss cone boundary into the hyperboloid. Thus, near the injection region $z'=0$, the mirror loss cone is partially filled for $v_\Vert <0$ due to the reflections by the electric field, Fig. \ref{fig:KineticR10_e}. For larger electron energies $E$ with $E>-e\phi_{tot}$, the magnetic mirror reflections dominate ($\phi_{tot}$ -- total potential drop). To the right of the maximum in the magnetic field for $z'>0.5$, there are three groups of electrons: a) the passing particles coming from the source and absorbed by the right wall; b) the particles coming from the source and reflected by the magnetic mirror and the electric field before they reach the wall; and c) the third group, the trapped particles that are separated from the source and can be reflected by the electric field from the right (including the collector sheath) and by the mirror force on the left. Particles can only enter this trapped region due to collisions (real or numerical) or non-stationary fluctuations such as transients in the initial stages \cite{SanchezPSST2018,Kim_2019}. The boundaries for the trapped particle in velocity space are obtained from the energy and momentum conservation:
\begin{equation}
    v_\perp = \sqrt{\left(\frac{2q}{m}(\phi_M - \phi(z^\prime)) - v_\parallel^2 \right) / \left(1 - \frac{B_M}{B(z^\prime)}\right)},
\label{Eq:MagCond}
\end{equation}
\begin{equation}
    v_\perp = \sqrt{\left(\frac{2q}{m}(\phi_W - \phi(z^\prime)) - v_\parallel^2 \right) / \left(1 - \frac{B_W}{B(z^\prime)}\right)}.
\label{Eq:PotCond}
\end{equation}

Here, the values of the potential and magnetic field are labeled by the index (M) and (W), for the mirror throat and the wall, respectively. The solid black line depicted in Fig. \ref{fig:KineticR10_e} represents the boundary from Eq.  (\ref{Eq:PotCond}) corresponding to the reflection by the electric field. The black dotted line describes the boundary from Eq.   (\ref{Eq:MagCond}) for trapping by the magnetic mirror. The electrons contained between both boundaries are fully trapped. In the regions near $z'\simeq 0.7$ and $z'\simeq 0.9$, one can observe populations of passing and reflected electrons. In our base case collisionless simulations, there are no reasons for electrons to enter the trapped region except for transient processes at the beginning of simulations and (perhaps) numerical PIC noise. We observe a relatively small number of trapped particles (less than 20\% of all particles in the expansion region) observed at $z'\simeq 0.7$ and $z'\simeq 0.9$  in Fig. \ref{fig:KineticR10_e}. 
We note that the effects of trapped 
particles are not expected to dramatically modify our results, e.g.  Ref. \onlinecite{ZhouPSST2021} shows the maximal increase of trapped electrons to 40\% in collisional case vs 25\%  in a collisionless situation similar to our result.  The role of collisions and trapped electrons are  further discussed in Section V.


\subsection{Finite ion temperature and expansion ratio effects}

\begin{figure}[h]
\centering
\captionsetup{justification=raggedright,singlelinecheck=false}
\includegraphics[width=0.85\textwidth]{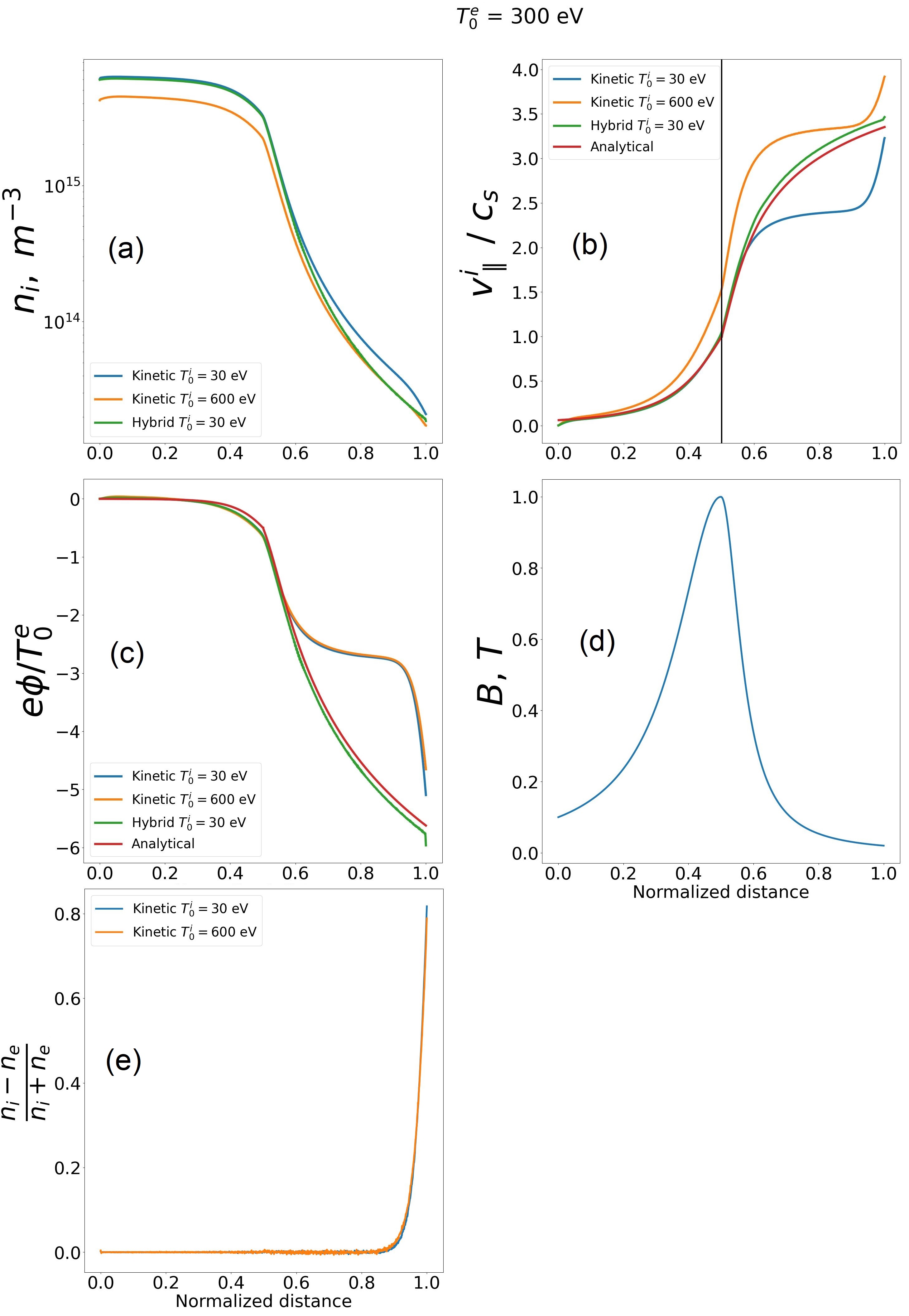}
\caption{Effects of the finite ion temperature in fully kinetic and hybrid models on Plasma density (a), ion velocity (b), potential profile (c), and deviation from quasineutrality (e), $T_0^e=300$ eV. The red lines in (b) and (c) show the results of analytical theory for cold ions, Eq. (\ref{m}).  The magnetic field profile with $R=10$, $K=50$ is shown in (d).}
\label{fig:Ion_effect}
\end{figure}

It was shown in Ref. \onlinecite{SmolyakovPoP2021} that in the quasineutral regime and under the assumption of isothermal electrons, the plasma velocity in the mirror configuration is uniquely defined by the magnetic field profile:
\begin{equation}
V_\Vert^i/c_s  =\left[ -W(-b^{2}\left( z\right) /e)\right] ^{1/2}.
\label{m}
\end{equation}
Here $W\left( y\right)$ is the Lambert function, which is the solution of the equation $W\exp \left( W\right) =y$, and $b\left( z\right) \equiv B\left( z\right) /B_{m}<1$, $B_{m}$ is the maximum value of the magnetic field, $e$ is Euler's number. The Lambert function has two branches in the real plane, $W_{0}\equiv W\left( 0,y\right)$ and $W_{-1}\equiv W\left( -1,y\right)$, which join smoothly at $W=-1$, for $y=-1/e$, corresponding to the sonic point at $b=1$. This analytical velocity profile is shown in Fig. \ref{fig:Ion_effect}b by the solid line in red. This analytical curve shows only a small difference with the results of our hybrid simulations with kinetic ions and Boltzmann isothermal electrons (shown in \ref{fig:Ion_effect}b in orange). The difference can be explained by a small but finite ion temperature in our simulations while the result in Eq. (\ref{m}) was obtained for cold ions. 

\begin{figure}[h]
\centering
\captionsetup{justification=raggedright,singlelinecheck=false}
\includegraphics[width=1\textwidth]{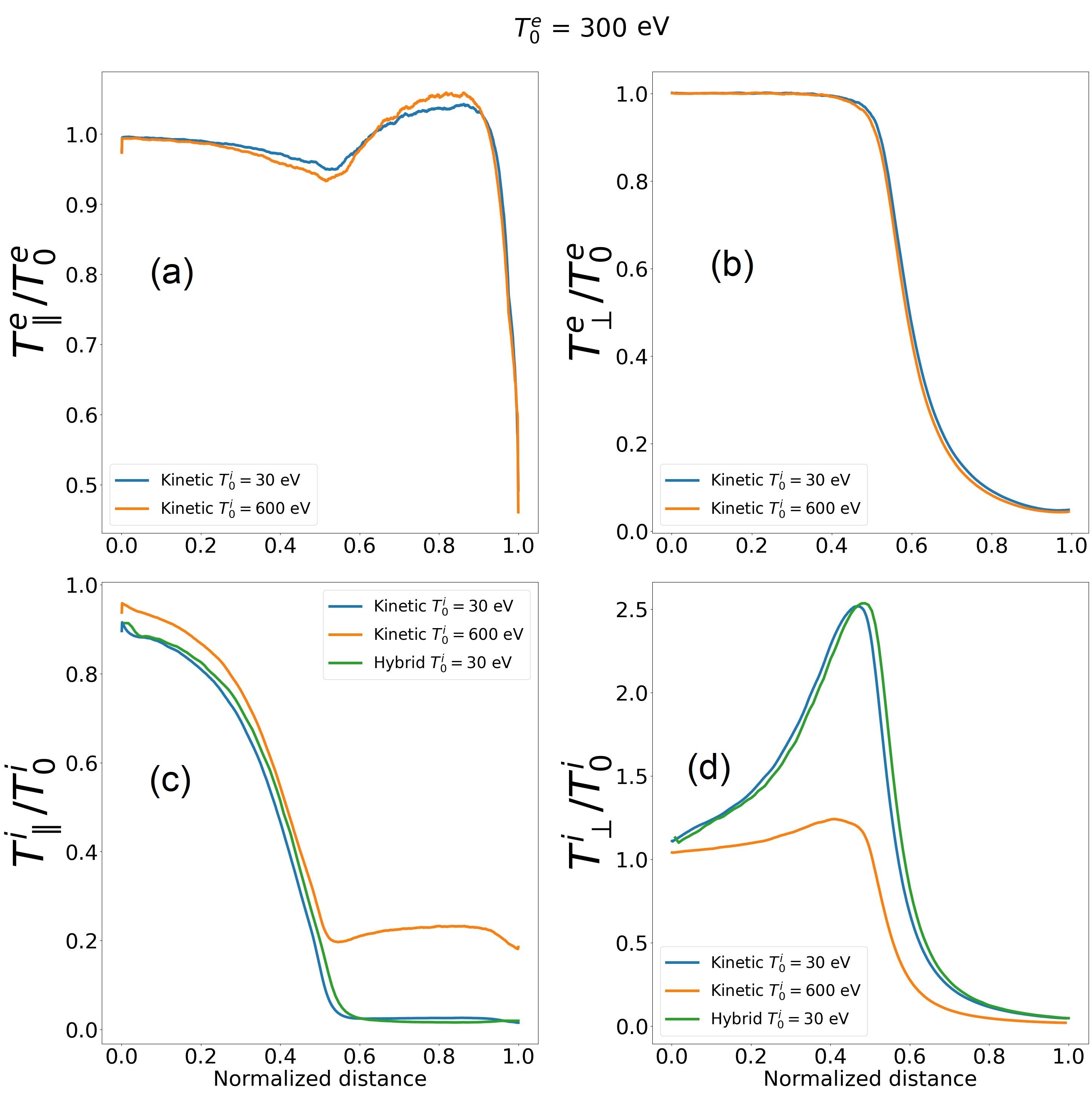}
\caption{Electron (a and b) and ion (c and d) temperature profiles for different temperatures of ions injected from the source, $T_0^i=30$ eV and $T_0^i=300$ eV, $T_0^e=300$ eV. Ion temperatures are shown for both fully kinetic and hybrid models.}
\label{fig:Ion_effect_temp}
\end{figure}

It is worth noting that our hybrid quasineutral model with $T^i_0 = 30 eV$ may experience numerical oscillations in the expansion region, leading to ion heating. The results in  Fig. \ref{fig:Ion_effect} and \ref{fig:Ion_effect_temp} were obtained with an increased number of particles per cell (PPC)  which eliminated the noise.  The mesh resolution was taken $0.01 \; m$, the timestep -- $2.6\times 10^{-10} \; s$, and the average PPC number was $10^4$.

The fluid theory with the two-pressure adiabatic model for ions predicts \cite{SaboPoP2022} that the finite ion pressure enhances plasma acceleration in the nozzle. This effect is related to the mirror force on ions with a finite perpendicular energy. This prediction is confirmed in our full kinetic simulations for two values of the ion temperature in the plasma source, 30 eV and 600 eV,  Fig.  \ref{fig:Ion_effect}b.

In the two-pressure CGL model, the plasma velocity at the sonic point is defined by the expression \cite{SaboPoP2022}: \begin{equation}
    v_s = \sqrt{\frac{T^e_\Vert + 3 T_\parallel^i}{m_i}},
    \label{vcs}
\end{equation}
where $T_\Vert^i$ is the ion parallel pressure at this point and $m_i$ is the ion/proton mass. The increase of the ion velocity at the maximum magnetic field at higher $T^i_0$ in our simulations can be seen in Fig.  \ref{fig:Ion_effect}b.  

\begin{figure}[h!]
\centering
\captionsetup{justification=raggedright,singlelinecheck=false}
\includegraphics[width=0.90\textwidth]{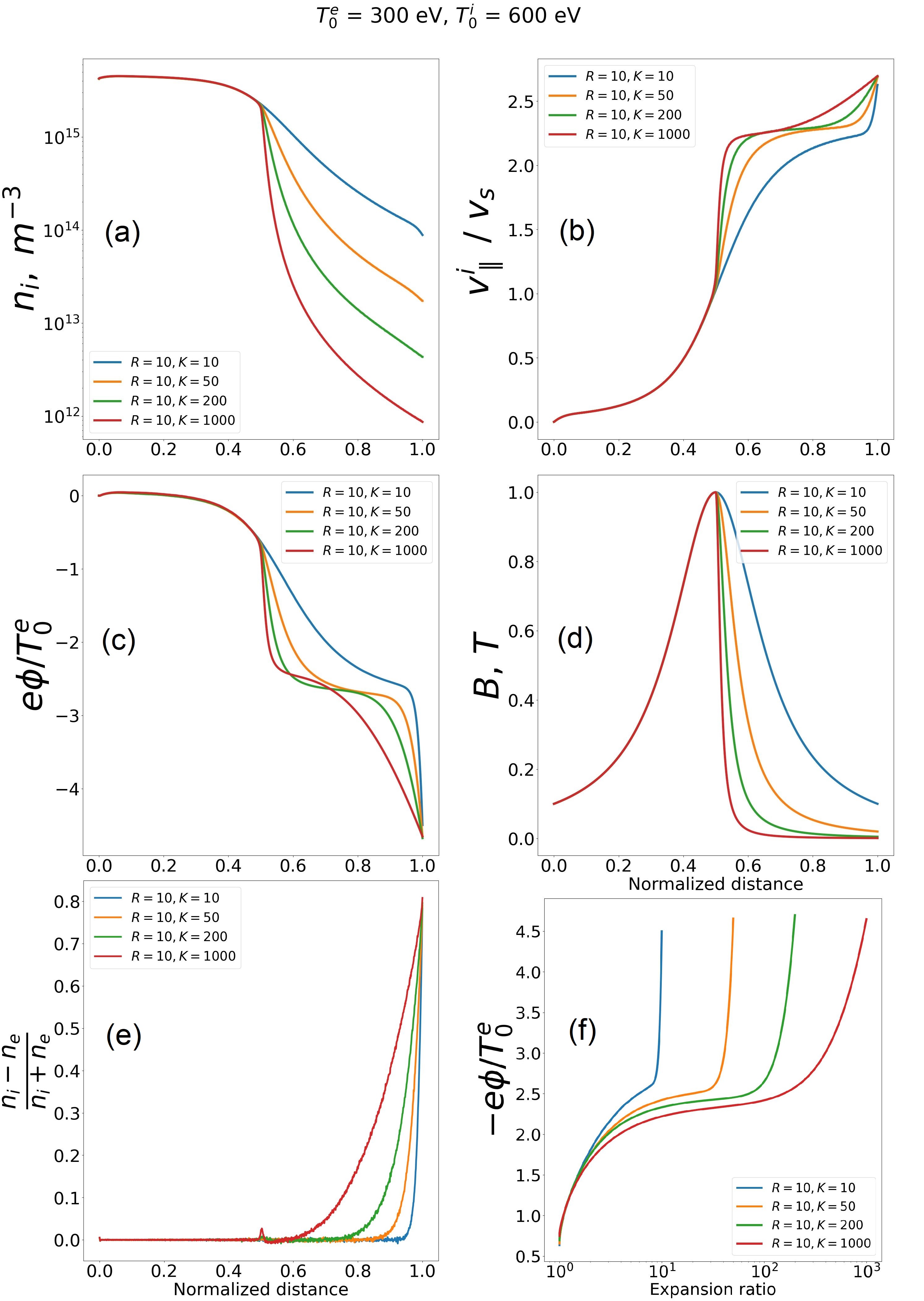}
\caption{  Plasma density (a), ion velocity (b), (c) potential, (d) magnetic field, and (e) deviation from quasineutrality  shown as functions of the normalized distance $z/L$ for different expansion ratios; (f) is the potential profile as a function of the expansion ratio;  the mirror length $L$  is kept the same for all $K$.}
\label{fig:Different_K}
\end{figure}

In our simulations, for $T_0^e=300$ eV and $T_0^i=600$ eV, at the location of the maximum magnetic field, we observe $T^i_\Vert \simeq 0.2 T^i_0$ and $T^e_\Vert \simeq 0.95 T^e_0$, from Figs. \ref{fig:Ion_effect_temp}a,b, and c. Then, in the units of the injection temperature for electrons, Eq. (\ref{vcs}) gives $v_s=\sqrt{2.15 T^e_0 / m_i}$, which is fairly close to the ion velocity observed in simulations, see Fig. \ref{fig:Ion_effect}b. 
One can note the jump in the velocity gradient at the sonic point, Fig. \ref{fig:Different_K}, due to the discontinuity of the second derivative of the magnetic field at the mirror throat. 

One has to note also that we do not expect the full agreement of the simulations with Eq. (\ref{vcs}), which was derived with the assumption of the constant and uniform (and isotropic) electron temperature and neglect of the ion heat fluxes. As one can see from Figs. \ref{fig:Ion_effect_temp}, the electron temperature is not constant and not isotropic. The parallel electron temperature is slowly decreasing in the converging part of the mirror (to the left of the mirror throat). There is a significant drop in the perpendicular electron temperature in the expander, Fig. \ref{fig:Ion_effect_temp}b. The parallel ion temperature for $T^i_0 = 600 \; eV$ rises in the expander, exhibiting a temperature increase similar in nature to that observed for electrons. 
Additionally, it should be noted that in full kinetic theory, taking heat fluxes into account, the definition of ion-sound velocity changes, and the location of the sonic point may differ slightly from that of the magnetic throat.



One can observe a reduction of plasma density at higher ion temperatures, Fig. \ref{fig:Ion_effect}. This can also be explained by the increase in plasma exhaust velocity, including the region near the plasma source (to the left of the magnetic throat). In our simulations (for both values of the ion temperature), the injected particle flux is the same. Therefore, the increased ion velocity results in a lower plasma density.

\begin{figure}[h!]
\centering
\captionsetup{justification=raggedright,singlelinecheck=false}
\includegraphics[width=0.8\textwidth]{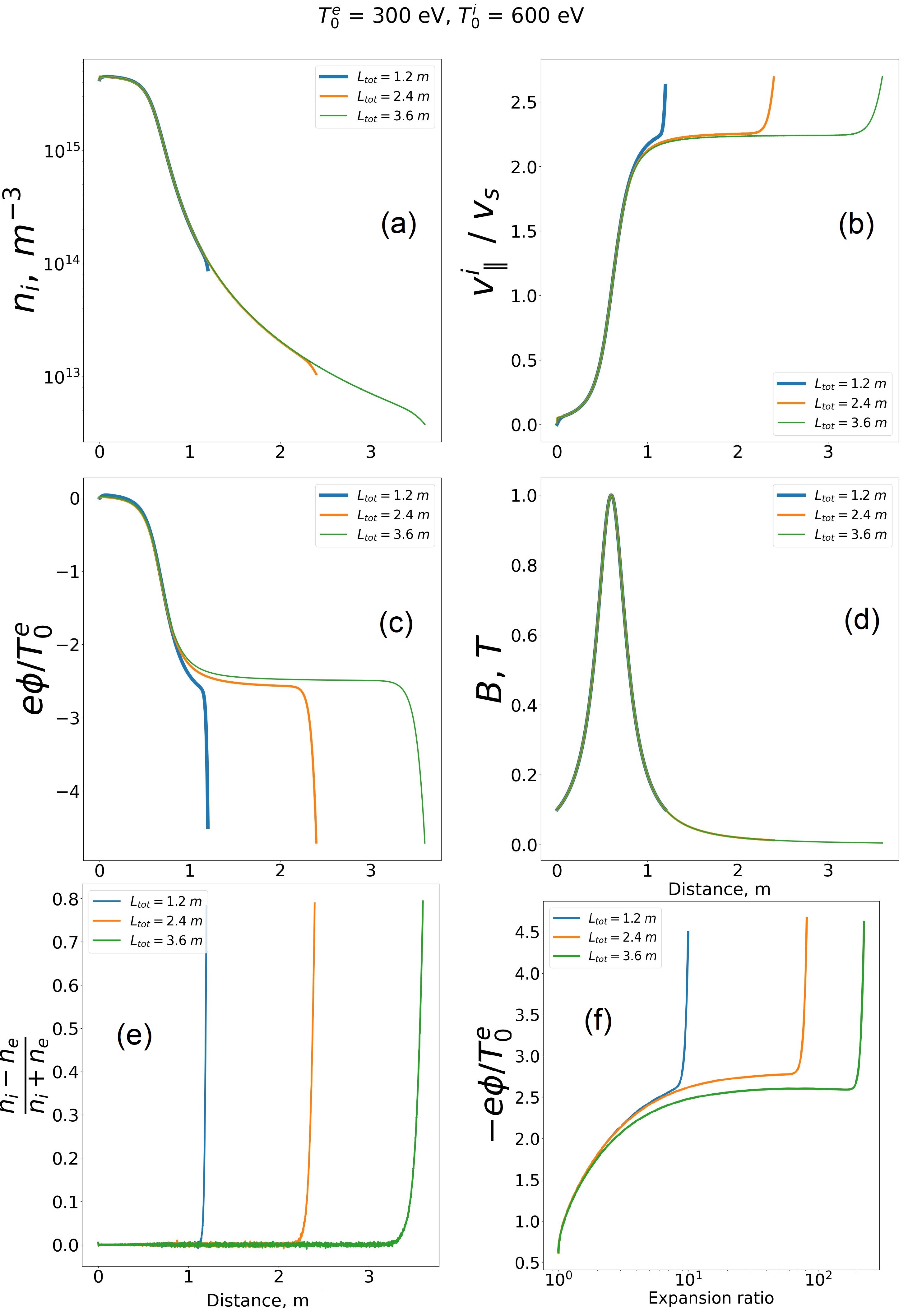}
\caption{Plasma density (a), ion velocity (b), potential (c) profiles in simulations with the magnetic field from Eq. (\ref{Eq:MagField}) and shown in (d). Deviation from quasineutrality (e) and potential profile as a function of the expansion ratio (f). Different colors show the results for different lengths of the expander with $z^\prime=z/L \geq 1$, $L = 1.2 \; m$; blue -- $L_{tot}=1.2\; m$ and $K = 10$, orange -- $L_{tot} = 2.4 \; m$ and $K = 82$, green -- $L_{tot} = 3.6 \; m$ and $K = 226$.}
\label{fig:extended_expander}
\end{figure}

It is interesting to look at the effect of the expansion ratio on the plasma flow in the mirror, keeping the same length of the mirror, as shown in Figs. \ref{fig:Different_K}. As expected, an increase in the expansion ratio results in the reduction of plasma density near the absorbing wall. Respectively, the sheath width is increasing. A notable effect is that the total potential drop across the whole mirror remains roughly constant, with the potential drop across the sheath and quasineutral regions also being very similar, as shown in Fig. \ref{fig:Different_K}c and e. Figures  \ref{fig:Different_K}d and f show how the potential profiles depend on the magnetic field.  One may note that in the expander region, the potential does not really follow the magnetic field even for the quasineutral region, i.e. the potential is different for different magnetic field profiles even though the values of the expansion ratio are the same, while the total potential drop across the quasineutral region stays roughly the same, \ref{fig:Different_K}f.  In these simulations with the constant mirror length, the increasing sheath width for large expansion becomes a large fraction of the total length of the expander as it is seen in  Fig. \ref{fig:Different_K}e for the deviation from quasineutrality and in the plasma velocity profiles, Fig. \ref{fig:Different_K}b.
 To further explore this behavior, we performed simulations by increasing the length of the expander, using Eq. (\ref{Eq:MagField}) for the magnetic field and extending the expander length to larger values of $z^\prime >1$, Fig. \ref{fig:extended_expander}. Figures \ref{fig:extended_expander}c and f show the potential profiles for different expander lengths $L_{tot}$. Now the length of the quasineutral region and non-quasineutral sheath are much better separated. One can observe from Figs. \ref{fig:extended_expander}a, c, e, and f that plasma density continues to drop and plasma expands quasineutrally without any further acceleration and keeping the same total potential drop across the whole quasineutral region. The same effect was reported in Ref. \onlinecite{SanchezPSST2018}. The total potential drop (and related plasma acceleration ) ceases to increase since the perpendicular particle energy (both in ions and electrons) is fully "used up" and there is no more free energy reservoir for the acceleration. The plasma velocity no longer increases, and plasma density continues to decrease following $n \sim B$, corresponding to the expansion at constant velocity. This expansion occurs quasineutrally, as shown in Fig. \ref{fig:extended_expander}e.




\subsection{Energy conservation  and axial  energy transport  in the mirror}
One of the crucial questions in the mirror confinement is energy losses from the confinement volume \cite{GotaNF2021}.
A critical question is electron losses which potentially can reduce electron temperature dramatically due to their high mobility. 
 A standard  figure of merit\cite{RyutovFST2005,SoldatkinaNF2020,SoldatkinaAIP2016} is the energy lost from the plasma source per one ion (equivalently per one ion-electron pair, since the flow is current-free), 
 \begin{equation}
 \eta=Q_{tot}/{T_0^e\Gamma},
 \end{equation} where $Q_{tot}=Q_{\Vert}^e+Q_{\Vert}^i$ 
 is the total energy flux from the source and $\Gamma$ is the particle flux. The particle and energy fluxes are defined as follows 
\begin{equation}
\Gamma_\parallel^{\alpha} = \int v_\parallel f^\alpha(r,t, \mathbf{v})d^3v, 
\end{equation}
\begin{equation}
Q_\parallel^{\alpha} = \frac{m}{2} \int v^2 v_\parallel f^\alpha(r,t, \mathbf{v}) d^3v,
\end{equation}
where $\alpha=(e,i)$, $\Gamma\equiv\Gamma_\Vert^e=\Gamma_\Vert^i$.

It is worth reminding that due to multiple reflections of electrons back to the source, energy taken from the source can be much higher than the electron temperature in the source, $\eta>1$.  It is reasonable to assume that the bulk of the electrons in the source have the Maxwellian distribution due to the energy confinement time (residence time in the source) being much larger than the collisional time.
It is important to note however that the distribution function coming from the source in the loss region of the phase space (loss cone with a cutoff due to the electric field)  is expected to be different from the assumed Maxwellian. The distribution function in this region is established due to the balance of the collisional diffusion in the velocity space and the axial streaming of the thermal electrons away. Since in practical applications, the collisional electron-electron frequencies are many orders of magnitude smaller than the bounce frequency, the distribution function in this region is strongly depleted compared to that of the Maxwellian distribution thus reducing the electron energy losses \cite{Pastukhov_1974,Sivukhin1965}. This is an important effect that has to be considered in the calculation of the actual axial losses from the magnetic mirror systems\cite{RyutovFST2005,KonkashbaevJETP1978}. A full analysis of the effects of collisions is beyond the scope of the present paper. Our reflux model for the plasma source makes an assumption that the loss cone is full: the electrons in the loss regions are randomly replenished from the original Maxwellian at every return to the source region. From the perspective of the energy losses from the confinement volume, it is the worst possible situation.  
Therefore our collisionless model with the Maxwellian reflux provides an upper bound for energy losses.  
The analysis of the electron and ion energy transport also confirms the energy conservation in the simulations and establishes the energy conservation properties for the collisionless base case. 

In our model, the conservation of density and energy for electrons and ions obtained from the drift-kinetic equation (\ref{dke}) are  given by the following equations
\begin{equation}
            \mathbf {B}\cdot \nabla \frac { \Gamma_{\parallel}^e}{B} = {\bf B\cdot} \nabla \frac { \Gamma_{\parallel}^i}{B}=0,\label{gc}
 \end{equation}
\begin{equation}
\mathbf{B\cdot }\nabla \frac{Q_{\Vert }^{\alpha }}{B}=q^{\alpha }n^{\alpha
}V_{\Vert }^{\alpha }E_{\Vert }.
\label{eca}
\end{equation}

Note that the energy fluxes, $Q_\parallel^{\alpha}$, are for the total energy in each species, i.e., include both thermal (random) energy and the kinetic energy of the directed flow. In general, the total energy flux can be written in the form
\begin{equation}
\mathbf{Q}=\left( \frac{mn}{2}\mathbf{V}^{2}+%
\frac{5}{2}p\right) \mathbf{V+\Pi \cdot V+q}.
\label{Q}
\end{equation}%
where $\mathbf{V}$ is the vector flow velocity , $\mathbf{q}$ is the heat flux vector, $p$ is the isotropic pressure, and $\mathbf{\Pi}$ is the viscosity tensor. Here, we only consider the parallel viscosity tensor responsible for the pressure anisotropy 
\begin{equation}
\mathbf{\Pi}=\left( p_{\Vert }-p_{\bot }\right) \left( \mathbf{bb-}\frac{1}{%
3}\mathbf{I}\right),
\end{equation}
where the random particle velocity $v^{\prime }_\Vert$ is defined as follows $v_\Vert =V_\Vert + v^{\prime }_\Vert$ and $V_\Vert$ is flow velocity.
Applying drift-kinetic approximation and neglecting the azimuthal components, Eq. (\ref{Q}) reduces for each species to 
\begin{equation}
\mathbf{Q=}Q_{\Vert }\mathbf{b}=\left( \frac{mn}{2}V_{\Vert }^{2}\mathbf{+}%
\frac{3}{2}p_{\Vert }+p_{\bot }\right) V_{\Vert }\mathbf{b}+\left( q_{\bot
}+q_{\Vert }\right) \mathbf{b},
\label{Qp}
\end{equation} 
where the heat fluxes are defined as follows 
\begin{equation}
q_{\Vert }=\frac{m}{2}\int {v^{\prime }_\Vert}^2 v^{\prime }_\Vert fd^{3}v,
\end{equation}%
\begin{equation}
q_{\bot }=\frac{m}{2}\int v_{\bot }^{2}v^{\prime }_\Vert fd^{3}v,
\end{equation}%

One can also write the total energy conservation in the form 
\begin{equation}
            {\bf B\cdot \nabla} \frac {Q^{e}_\parallel +Q^{i}_\parallel} {B} =  { J_\parallel} E_{\parallel}.
            \label{ec}
 \end{equation}
 Here $J_\parallel=J_{\parallel}^i+J_{\parallel}^e$ is the total current, and the term on the right-hand side describes the energy exchange between thermal plasma energy and the electromagnetic field.
  
Equations (\ref{eca}) describe the energy exchange between electrons and ions mediated by the electric field. These equations can be integrated, so that the total potential drop across the nozzle can be related to the energy and particle fluxes.     
\begin{equation}
\left( \frac{Q_{\Vert }^{i}}{B}\right) _{z^\prime=1}\ -\left( \frac{Q_{\Vert }^{i}}{%
B}\right) _{z^\prime=0}\mathbf{=}-e\frac{\Gamma _{\Vert }^{i}}{B}\left( \phi
_{z^\prime=1}-\phi _{z^\prime=0}\right),\label{pi}
\end{equation}%
\begin{equation}
\left( \frac{Q_{\Vert }^{e}}{B}\right) _{z^\prime=1}\ -\left( \frac{Q_{\Vert }^{e}}{%
B}\right) _{z^\prime=0}\mathbf{=}e\frac{\Gamma _{\Vert }^{e}}{B}\left( \phi
_{z^\prime=1}-\phi _{z^\prime=0}\right).\label{pe}
\end{equation}

It is convenient to normalize the magnetic field to $B_0$ and rewrite equations (\ref{gc}) and (\ref{ec}) as 
 \begin{equation}
            \mathbf {\Hat{B}}\cdot \nabla  \frac{\Gamma_{\parallel}^e}{\Hat{B}} = \mathbf {\Hat{B}}\cdot \nabla \frac{\Gamma_{\parallel}^i}{\Hat{B}}=0,
 \end{equation} 
\begin{equation}
            \mathbf {\Hat{B}}\cdot \nabla \frac {Q^{e}_\parallel +Q^{i}_\parallel} {\Hat{B}} =  { J_\parallel} E_{\parallel}. \label{ec2}
 \end{equation}

\begin{figure}[h]
\centering
\captionsetup{justification=raggedright,singlelinecheck=false}
\includegraphics[width=1\textwidth]{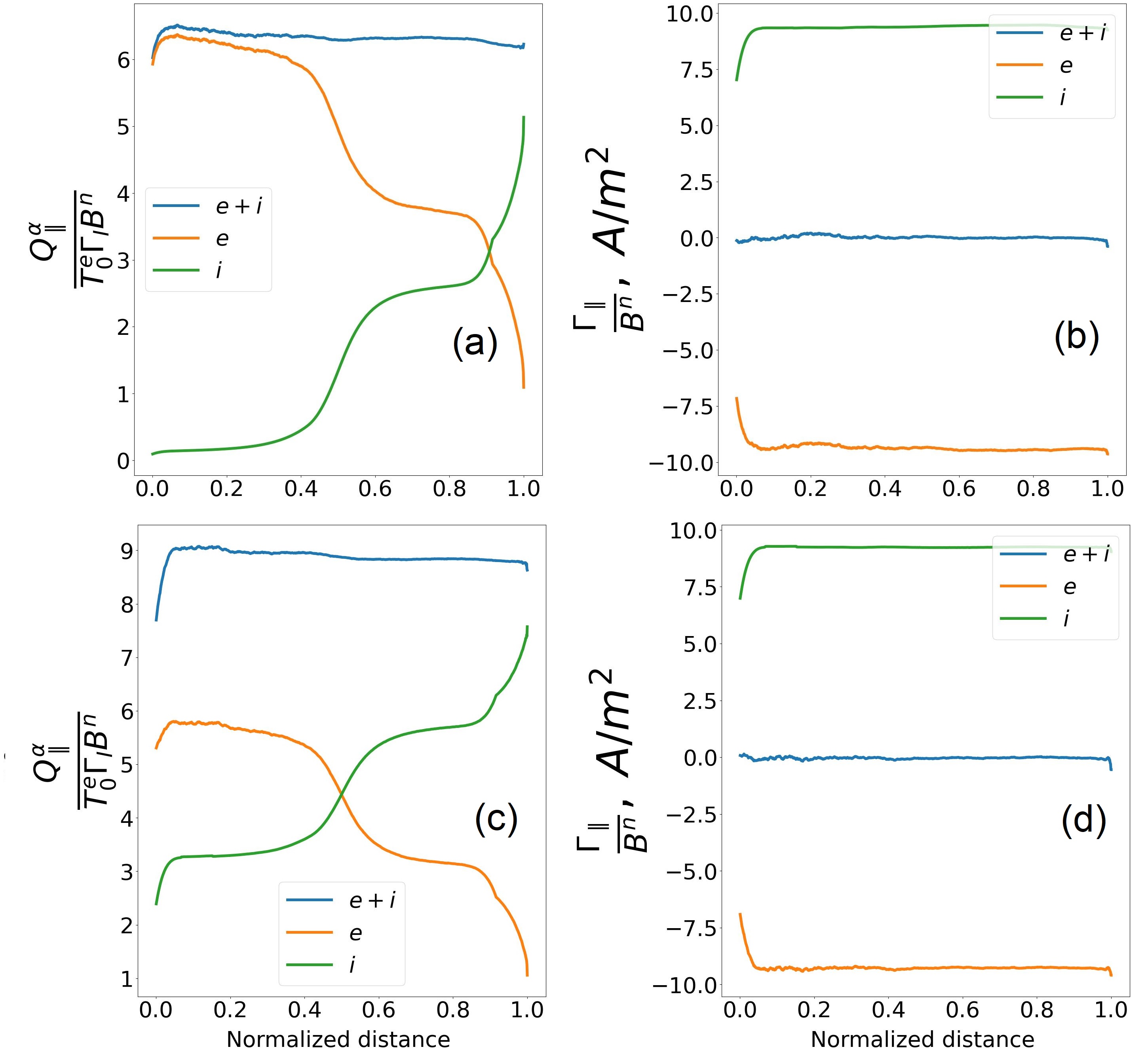}
\caption{(a) Axial energy transport and energy conservation, and (b) current conservation for $T_0^e = 300$ eV, $T_0^i = 30$ eV; (c) axial energy transport and energy conservation, and (d) current conservation for $T_0^e = 300$ eV, $T_0^i = 600$ eV; $R=10, K=50$.}
\label{fig:conserv}
\end{figure}

Normalized fluxes of particles and energy are shown in Figs. \ref{fig:conserv}. Good conservation of the electron and ion current in our simulations can be seen in Figs. \ref{fig:conserv}b and d.  This figure also shows that the total current is zero, $J_\Vert=0$, consistent with the equal injection rates for electrons and ions from the plasma source at the left boundary.

Equation (\ref{Qp}) shows that the total energy flux consists of the convective flux of the thermal (random) energy, the convective flux of the kinetic energy of the directed flow, and the heat fluxes. The ion and electron energy fluxes are shown in Figs. \ref{fig:conserv}a and c. The ions get their energy from the initial injection and from the accelerating electric field. The electron energy flux injected at the source gradually decreases due to the reflections of electrons by the mirror and electric forces. At the right boundary, the electron energy flux is determined by a small fraction of the electrons that overcome the total potential barrier in the Yushmanov potential and reach the wall.

The electrons lose their energy in reflections by transferring it to the electric field, which subsequently accelerates ions. A substantial exchange between electrons and ions mediated by the electric field is illustrated in Figs. \ref{fig:conserv}a and c.  The total energy exchange between plasma and the electrostatic field is described by the last term in Eq.(\ref{ec}).  
 For the current-less situation $J_\parallel=0$, the total energy in the electric field remains constant so that the sum of the electron and ion energy fluxes remains constant. Note that a momentum-conserving scheme is used in these simulations. We observe a weak variation of the total energy flux in the converging part of the mirror where plasma density is high. This numerical error is further discussed in Section V.

Of particular interest is the value of the total energy flux representing the losses to the wall. For our base case parameters, $T_0^i=30\;eV <T_0^e=300 \; eV$, the initial ion energy (at the left boundary) is small and can be neglected, as seen in Fig. \ref{fig:conserv}a. The total energy flux from the plasma source is mostly in the electron component and
as shown in Fig. \ref{fig:conserv}a, this flux to the wall per one electron-ion pair is of the order of $6T_0^e$. Near the absorbing wall, the total energy flux (per one electron-ion pair) is the sum of the kinetic energy of the electrons overcoming the potential barrier and reaching the wall and the kinetic energy of the ions. The latter part is equal to the change in the electric potential of the electrons reaching the wall if the ion energy at the injection is neglected. In units of the electron injection temperature, the energy loss can be expressed as

\begin{equation}
\eta =  \frac{1}{T_0^e} \left[\left( - e\phi_{tot} + W^e_{wall} \right) \right],
\label{eta}
\end{equation}
where $\phi_{tot}$ is the total potential drop, and $W^e_{wall} = \sum_j^N m_e v^2_j / 2 N$ (with $N$ being the total number of absorbed particles per time step).
In our simulations,  the potential drop   $-e\phi_{tot}\simeq 5 T_0^e$  and the direct calculations gives  $W^e_{wall}\simeq T_0^e$ so that  $\eta \simeq 6$ which is consistent with the  $(Q_\Vert^e+Q_\Vert^i)/(\Gamma_\Vert T_0^e) \simeq 6$ as illustrated in Fig. \ref{fig:conserv}a.

For the case of higher ion temperature, $T_0^i=600$ eV, the total energy loss for an ion-electron pair is around $9T^e_0$, where the order of $3T^e_0$ ($\approx 3T^i_0/2 $) ions get from the injection, $5T^e_0$ is transferred from the electrons, and $T^e_0$ is carried by the over-barrier electrons, Fig. \ref{fig:conserv}c.   The latter value is about the same energy as for the case of the lower ion energy. Since, in the current-less case, the total energy flux is constant in the axial direction, the net energy flux to the wall is a sum of energy fluxes in electrons and ions at $z^\prime=0$. These fluxes can be calculated from the anisotropic distribution functions at $z^\prime=0$, as shown in Figs. \ref{fig:KineticR10_e} and   \ref{fig:KineticR10_i}. The shape of the electron and ion distribution at this location is determined by the loss cone which, for each species, is determined by the combination of the magnetic mirror and electric field reflections. It is worth noting that the injected flux Maxwellian distributions carry very large energy fluxes, e.g., of the order of $Q_\Vert \simeq n v_{t\alpha} T_\alpha$ for each species. However, a large fraction of injected particles are confined, reflected back to the wall, and then replaced by random particles from the distribution with an original injection temperature, so the net energy flux of particles inside the loss cone is much lower.

\begin{figure}[htp]
\centering
\captionsetup{justification=raggedright,singlelinecheck=false}
\includegraphics[width=0.5\textwidth]{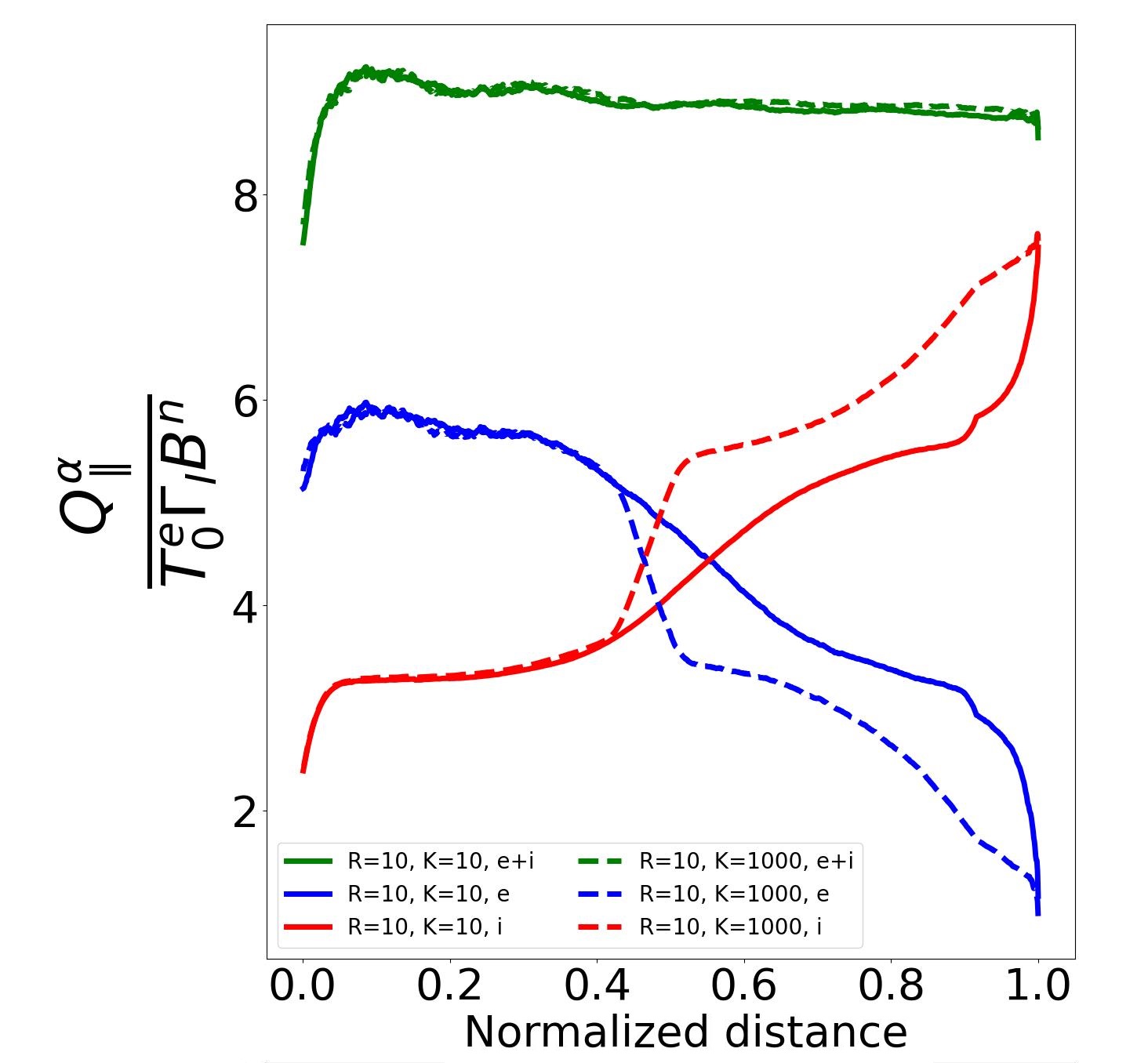}
\caption{Axial energy transport and energy conservation for different expansion ratios: $R=10$, $K=10$ and $R=10$, $K=1000$, $T_0^e = 300$ eV, $T_0^i = 600$ eV.}
\label{fig:conserv_10_1000}
\end{figure}

For all our simulations the energy loss factor can be directly related to the total potential drop as $\eta \approx (-e\phi_{tot} + T^e_0 + 3T^i_0 / 2) / T^e_0$, e.g.  Fig. \ref{fig:conserv_10_1000} for the same parameters as in Fig. \ref{fig:Different_K} shows that $\eta$ for the  $K$ values between $K=10$ and $K=1000$ are roughly the same.   


\subsection{Ion heat fluxes  in full kinetic, hybrid, and extended hydrodynamic models simulations}

\begin{figure}[h]
\centering
\captionsetup{justification=raggedright,singlelinecheck=false}
\includegraphics[width=0.9\textwidth]{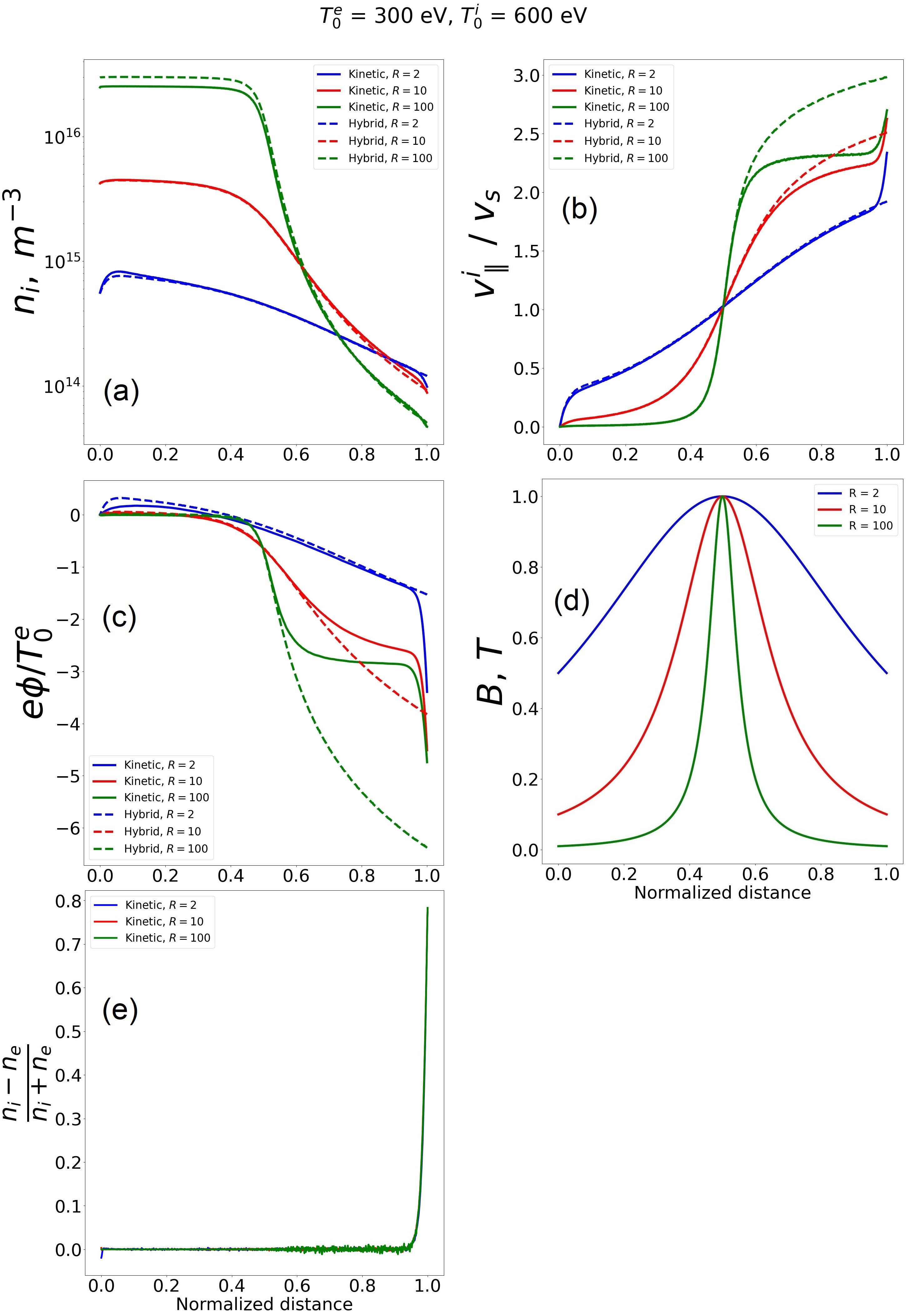}
\caption{Plasma density (a), ion velocity (b), potential (c), and deviation from quasineutrality (e) profiles as a function of the mirror ratio (d) in full kinetic and hybrid models, symmetric mirror $R=K$.}
\label{fig:Kin_boltz_conc}
\end{figure}

\begin{figure}[h]
\centering
\captionsetup{justification=raggedright,singlelinecheck=false}
\includegraphics[width=1\textwidth]{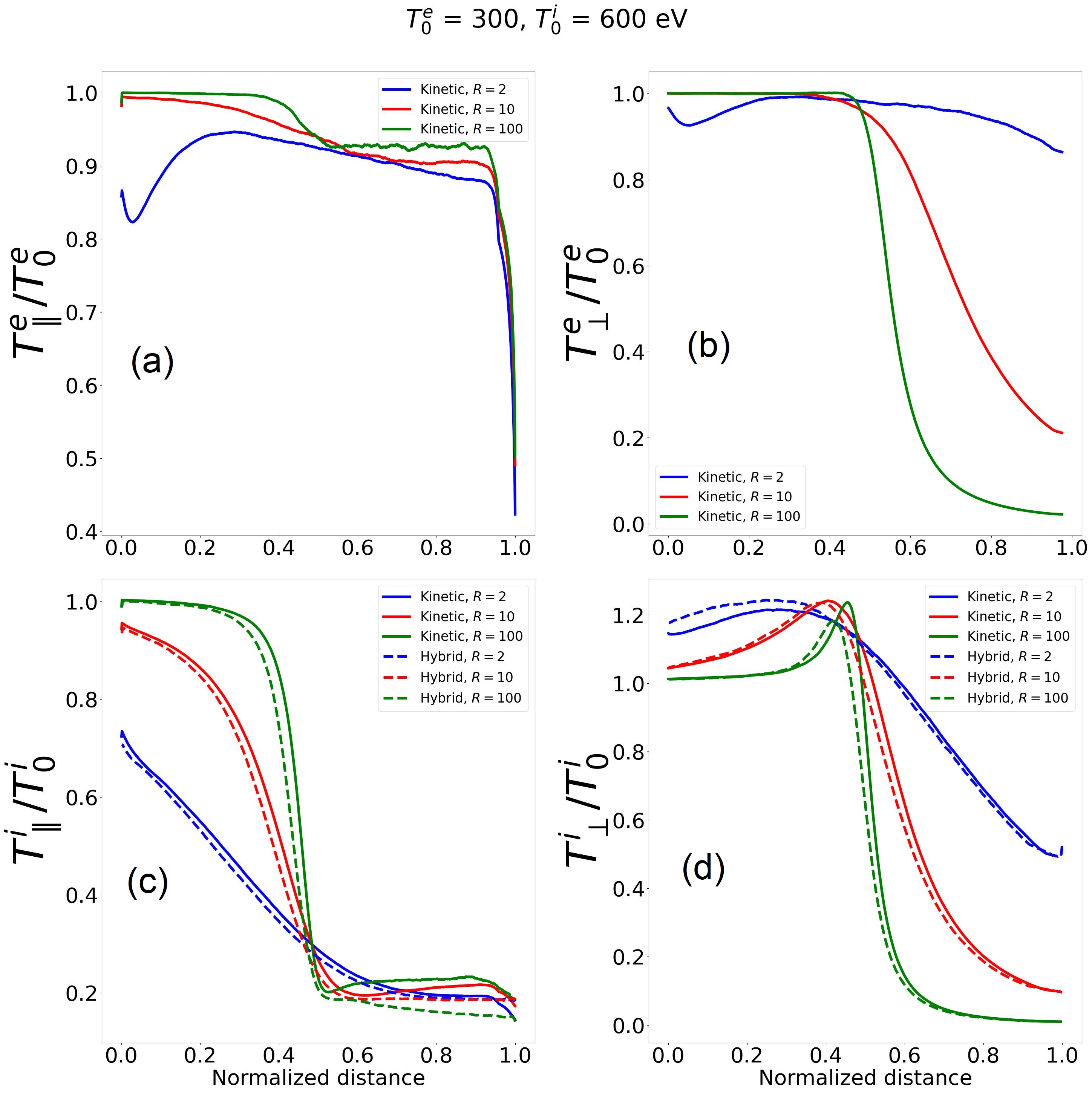}
\caption{Electron (a and b) and ion temperature (c and d) profiles as a function of the mirror ratio in full kinetic and hybrid models, symmetric mirror $R=K$.}
\label{fig:Kinc_Boltz_temp}
\end{figure}

\begin{figure}[h]
\centering
\captionsetup{justification=raggedright,singlelinecheck=false}
\includegraphics[width=1\textwidth]{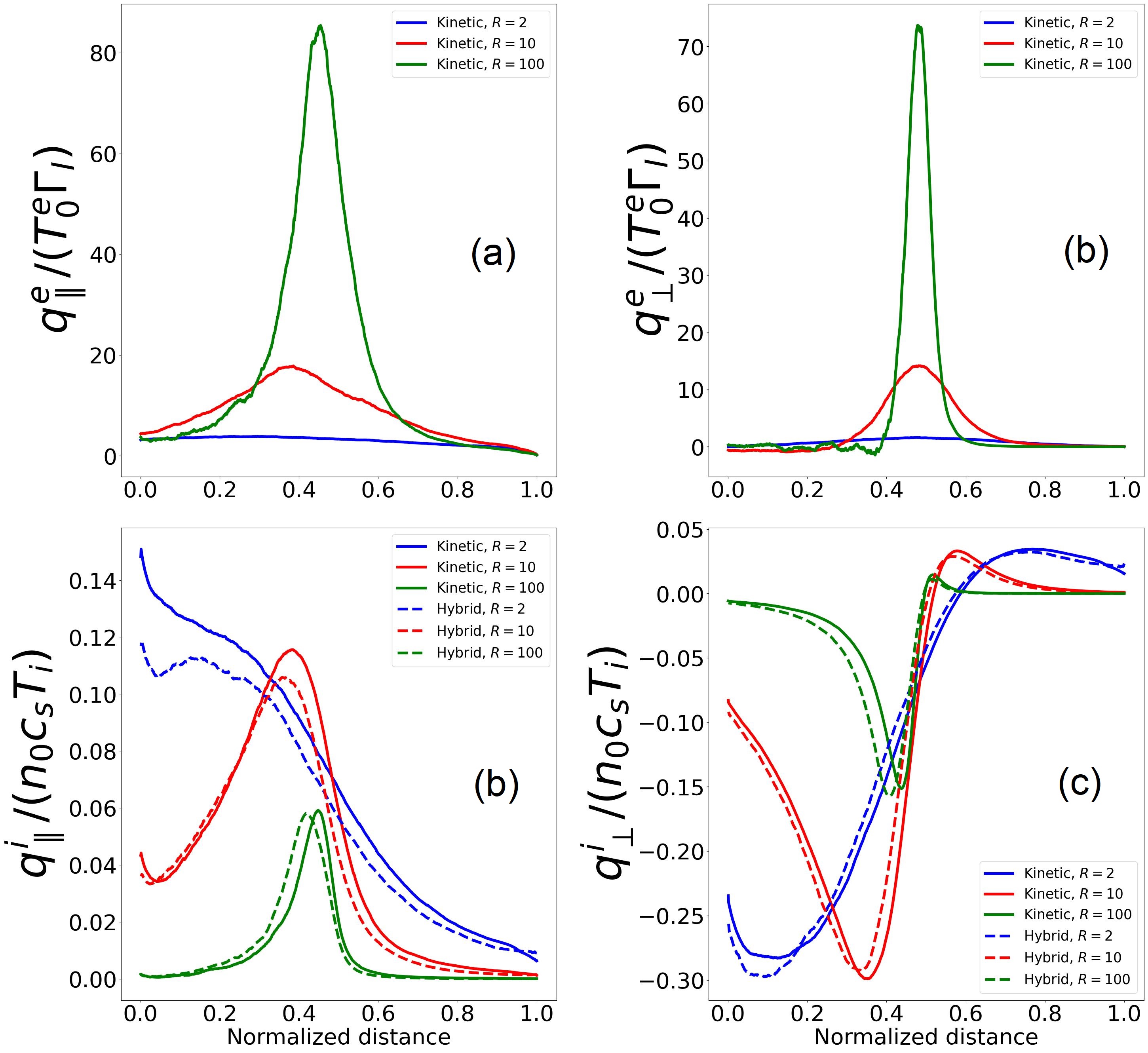}
\caption{Electron (a and b) and ion (c and d) heat fluxes in full kinetic and hybrid models, symmetric mirror $R=K$. $\Gamma_I$ is the electron flux at the left wall.}
\label{fig:Kinc_Boltz_heat}
\end{figure}

 An assumption of isothermal electrons is often used in simple theories of plasma flow in the magnetic mirror.   It is of interest to compare the results of the full kinetic calculations with the hybrid quasineutral model with isothermal electrons. This comparison is shown in Figs.  \ref{fig:Kin_boltz_conc}, \ref{fig:Kinc_Boltz_temp}, and \ref{fig:Kinc_Boltz_heat} for $T_0^e=300$ eV and $T_0^i=600$ eV. Obviously, the quasineutral model does not reproduce the sheath physics. The plasma potential in the full kinetic model closely follows the potential from the hybrid model in the converging part of the mirror.   In the expanding part, the potential and ion velocity profiles start to diverge strongly, especially for large expansion ratios. In part, the difference occurs because of the influence of the sheath structure. Another, probably more important, reason is the electron cooling and the anisotropy of the electron pressure in the expander region. In the converging part, the perpendicular electron energy remains pretty much constant,  Fig. \ref{fig:Kinc_Boltz_temp}b, due to a large number of electrons reflected by the mirror. The reduction of the parallel electron energy in the converging part is not significant,  Fig. \ref{fig:Kinc_Boltz_temp}b, so the electron temperature in this region may be considered approximately isotropic and isothermal, similarly as in the hybrid model. In Ref. \onlinecite{MartinezSPoP2015} authors observed the same behavior. As a result, all plasma parameters (density, ion velocity, and potential) in the converging part of the mirror in the hybrid and kinetic models are very similar, as seen in Figs. \ref{fig:Kin_boltz_conc}a, b, and c. In the expander region, the electron perpendicular temperature drops sharply, Fig. \ref{fig:Kinc_Boltz_temp}b, since many of the electrons with large values of the perpendicular energies are reflected before the mirror maximum. The parallel electron temperature also decreases, Fig. \ref{fig:Kinc_Boltz_temp}a.  Therefore, for large expansion ratios, electron cooling that is fully included in the kinetic model is a dominant mechanism of the difference between the kinetic and hybrid models. One can see from Fig. \ref{fig:Kin_boltz_conc}b, c, and e that the potential and ion velocity profiles start to diverge well before the deviations from the quasineutrality become noticeable. A similar effect is also observed in Fig. \ref{fig:Ion_effect}c and e.

The main differences between the hybrid and full kinetic models are in the potential and ion velocity profiles, Figs. \ref{fig:Kin_boltz_conc}b and c. The total potential drop and ion velocity in the hybrid model increase with the expansion ratio, $\delta \phi \simeq T_e \ln K$. In the hybrid model with the isothermal electrons, the energy transfer from electrons to ions remains large (formally infinite) in the expander region, resulting in the higher kinetic energy of the accelerated ions, diverging logarithmically with $K$, Fig. \ref{fig:Kin_boltz_conc}b. In the full kinetic model, the changes in the parallel and perpendicular electron pressure are related to the compression/expansion work on electrons, the work of the electric fields, and heat fluxes, as seen in Eq. (\ref{Qp}) for the total energy flux. In our injection and reflux model,  the energy transferred from the electrons to ions is fixed by the amount of energy injected into the electron component. As a result, the total potential drop (and the final ion energy) quickly saturates for large $K$ and becomes independent of $K$, see Figs. \ref{fig:Implicit}c and b.

The ion temperature and ion heat fluxes (both perpendicular and parallel) are very similar in the hybrid and kinetic models, as seen in Figs. \ref{fig:Kinc_Boltz_temp}c and d  Figs. \ref{fig:Kinc_Boltz_heat}a and b.             
The ion temperature effects on plasma flow in the mirror configurations were studied previously with two-pressure anisotropic ion pressure models in Ref. \onlinecite{SaboPoP2022, TogoCPP2018, TogoNF2019}. The simplest model \cite{SaboPoP2022} is based on the two-pressure adiabatic CGL model \cite{CGL1956}, which assumes that the heat fluxes are zero, which is, in general, not justified. In weakly collisional plasmas relevant to many applications, the closures for the heat fluxes are difficult \cite{SnyderPoP1997}. Different closures for the ion heat flux due to collisional, ad-hoc free streaming corrections, wave-particle interactions, and anomalous transport contributions were proposed \cite{SnyderPoP1997, ZawaidehPF1988, ZawaidehPF1986} and implemented to model plasma flow in the Scrape-Off Layer (SOL) and flux-expanding divertors \cite{TogoCPP2018, TogoNF2019, ZhaoCPP2016, FundamenskiPPCF2005}. The role of passing and trapped particles and ambipolar potential on the heat fluxes along open magnetic field lines was discussed in Ref. \onlinecite{LePoP2010, GuoPoP2012}. 

\begin{figure}[h!]
\centering
\captionsetup{justification=raggedright,singlelinecheck=false}
\includegraphics[width=1\textwidth]{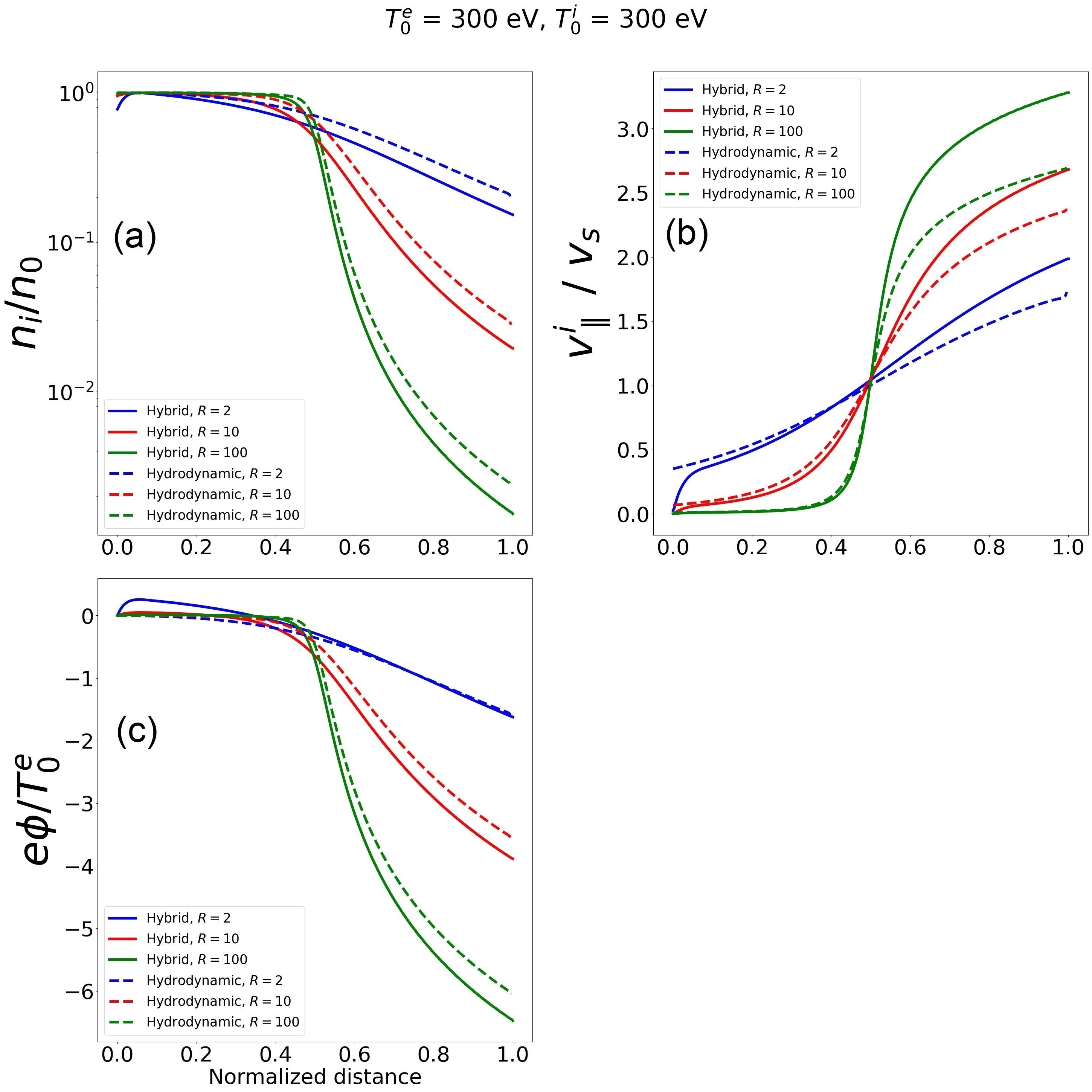}
\caption{Comparison of plasma density (a), ion velocity (b), and electrostatic potential (c) profiles in the extended hydrodynamic and hybrid (kinetic ions + Boltzmann electrons) models. Symmetric magnetic field profiles with $R=K$ are used (Fig. \ref{fig:Kin_boltz_conc}d), with $T_0^e=300$ eV, $T_0^i=300$ eV.}
\label{fig:HydroConc}
\end{figure}

\begin{figure}[h!]
\centering
\captionsetup{justification=raggedright,singlelinecheck=false}
\includegraphics[width=1\textwidth]{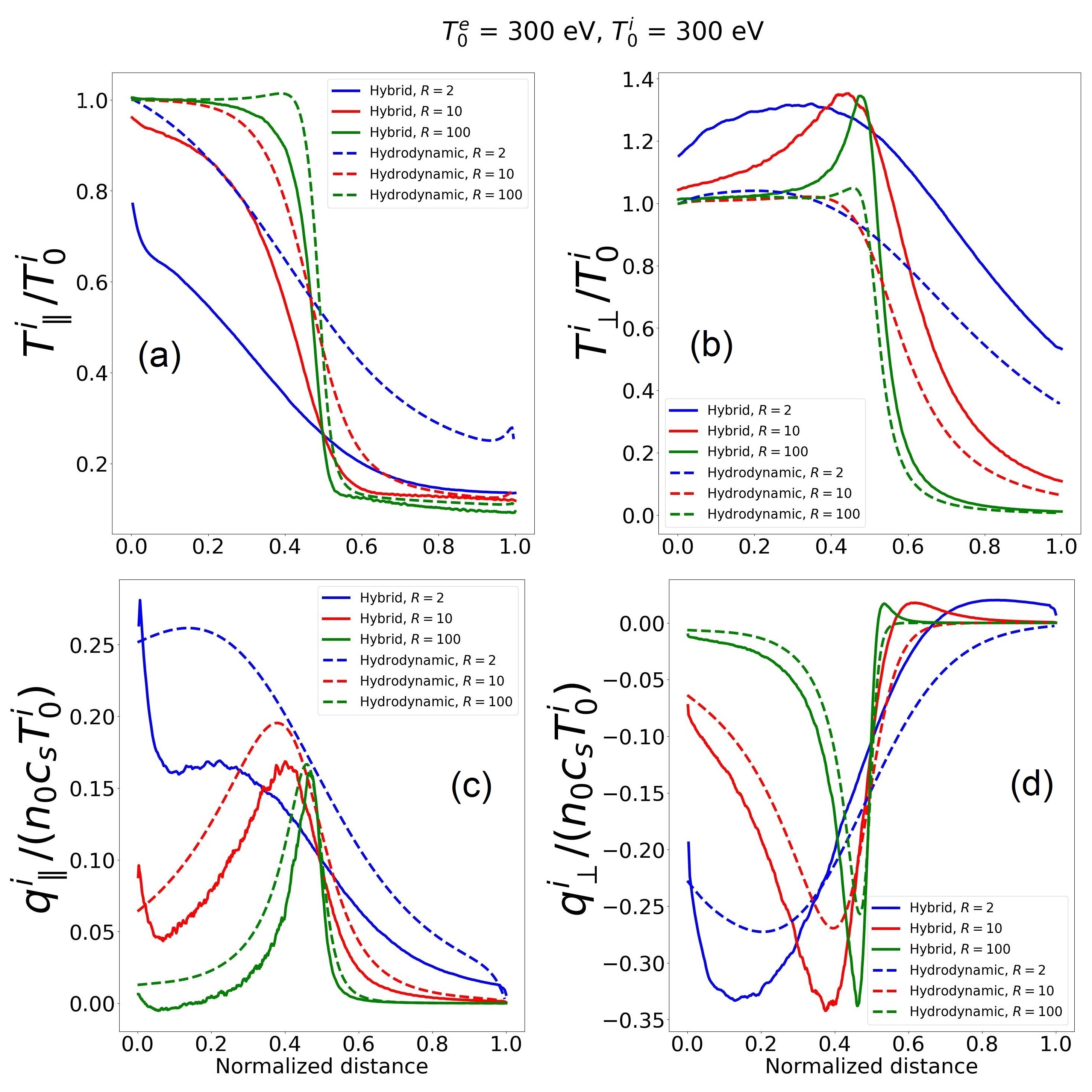}
\caption{Comparison of ion temperature (a) for parallel and (b) for perpendicular directions and ion heat fluxes (c) for parallel and (d) for perpendicular directions obtained in the extended hydrodynamic and hybrid (kinetic ions + Boltzmann electrons) models, symmetric magnetic field profiles with $R=K$ are used, with $T_0^e=300$ eV, $T_0^i=300$ eV.}
\label{fig:HydroTemp}
\end{figure}

We have previously used an extended hydrodynamic model for ions to model collisionless plasma flow in the mirror field, taking into account finite ion temperature effects \cite{Sabo2}. In this model, the heat fluxes $q_{\Vert i}$ and $q_{\perp i}$ are included in the evolution of the ion parallel and perpendicular pressures. The heat fluxes themselves are calculated from the time-dependent evolution equations for $q_{\Vert i}$ and $q_{\perp i}$. These equations are closed by the assumption that all fourth-order moments are calculated by using a two-temperature Maxwellian distribution. For completeness, the full system of extended hydrodynamic equations for ions is given in Appendix C. In Ref. \onlinecite{Sabo2}, characteristics of plasma acceleration were obtained using extended fluid equations for ions, Boltzmann approximation for electron density, and the quasineutrality assumption, as in the hybrid model.  
In Figs. \ref{fig:HydroConc} and \ref{fig:HydroTemp} we compare the results from Ref. \onlinecite{Sabo2} with our hybrid model which uses the same Boltzmann approximation for electrons and hydrodynamic solution for ions with $T_0^i=T_0^e=300$ eV and symmetric mirror $R=K$.
It is interesting to note that for a high mirror ratio $R=100$, the general behavior of the ion temperature and heat flux are similar in the extended fluid model and hybrid calculations, Figs. \ref{fig:HydroConc} and \ref{fig:HydroTemp}. The parallel ion temperature and parallel heat fluxes are in good agreement in both models, Fig. \ref{fig:HydroTemp}a and c. The perpendicular ion temperature and perpendicular heat fluxes are somewhat different. The notable difference is observed for the ion velocity profile, Fig. \ref{fig:HydroConc}b, while the plasma density and potential are fairly close, Fig. \ref{fig:HydroConc}a and c. It appears that the difference in the ion velocity originates from the large perpendicular ion temperature (in the region near the maximum magnetic field) in the fluid model, Fig. \ref{fig:HydroConc}b.  The mirror force due to the large perpendicular ion velocity (temperature) provides additional ion acceleration in the expander region resulting in higher final ion velocity.

\section{EDIPIC simulations of high-density regimes  with an implicit algorithm}

In this work, we use WarpX and EDIPIC (with modifications as described above in Section II.C) to compare their performance and determine the practical limits for the simulations with both codes. Explicit PIC simulations of realistic dimensions (a few meters in length) and densities of the order of $10^{18}$ $m^{-3}$ and larger are resource-intensive and may become impractical. While the current release of WarpX is explicit, EDIPIC \cite{SydorenkoTh2006} employs the implicit algorithm \cite{Gibbons1995TheDD}. Therefore, it is of interest to extend the study into the high-density regimes, exploring the advantages of the implicit EDIPIC code.

\begin{figure}[htp]
\centering
\captionsetup{justification=raggedright,singlelinecheck=false}
\includegraphics[width=1\textwidth]{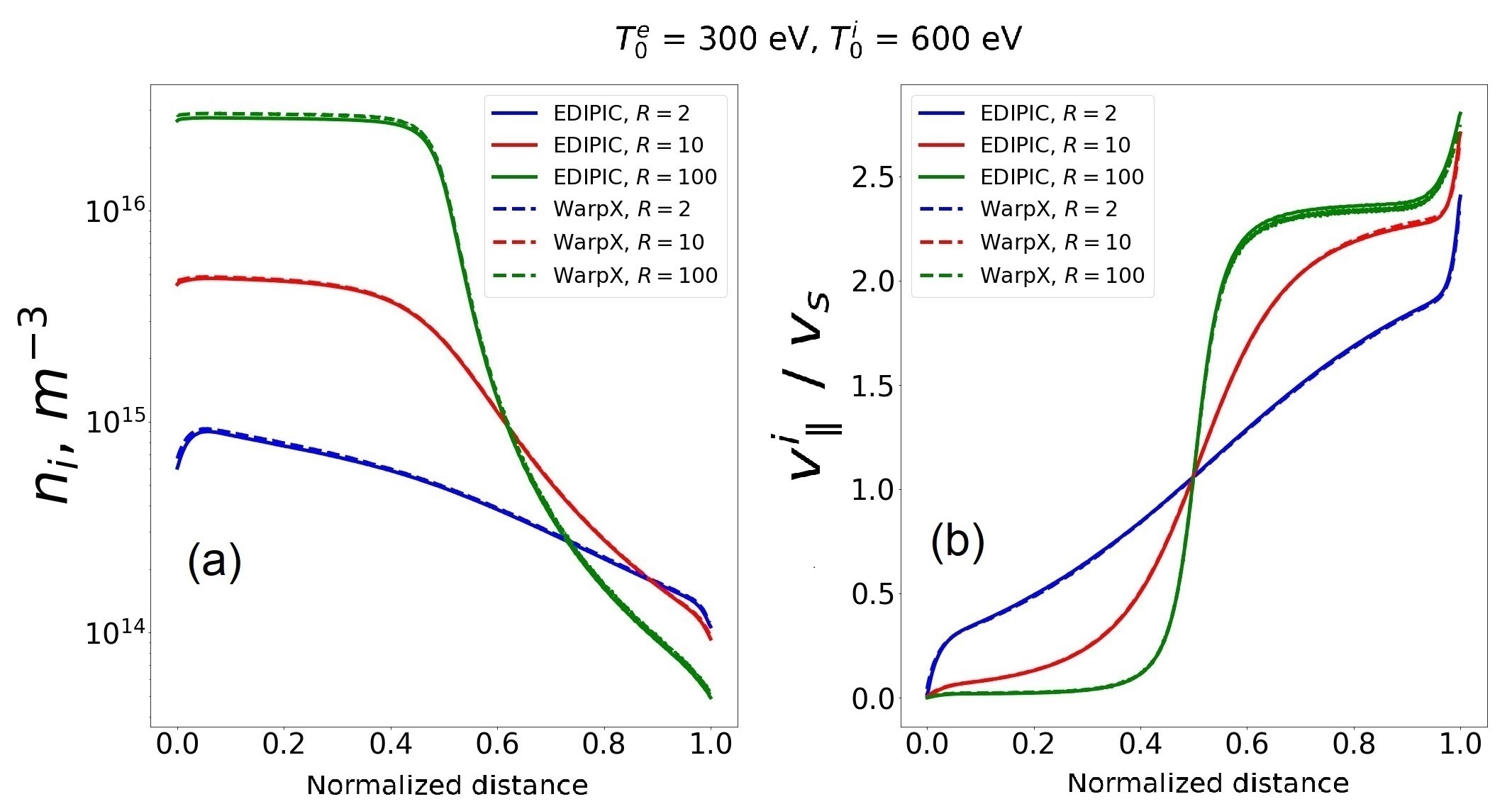}
\caption{EDIPIC and WarpX results for different $R$ values of the symmetric mirror, $T^e_0 = 300 \; eV$, $T^i_0 = 600 \; eV$ for ion concentration (a) and velocity (b). The WarpX simulations for $R=2$ and $R=10$ and all EDIPIC runs were performed with $\Delta t = 5 \times 10^{-11} \; s^{-1}$ and $\Delta z = 2 \times 10^{-3} \; m$. The time and spatial resolution for WarpX with $R=100$ had to be increased four times to achieve the agreement with EDIPIC.}
\label{fig:WarpXvsEDIPIC}
\end{figure}

To compare the WarpX and EDIPIC results, we have performed the simulation of the symmetric mirror $R=K$ with different mirror ratios for $T^e_0 = 300 \; eV$, $T^i_0 = 600 \; eV$. These simulations are performed with $\Delta t = 5 \times 10^{-11} \; s^{-1}$ and $\Delta z = 2 \times 10^{-3} \; m$, which are two times larger compared to those for the base case. The simulation results for $R=2$ and $R=10$ perfectly agree as seen in Fig. \ref{fig:WarpXvsEDIPIC}. However, the results for $R=100$ for WarpX (not shown in  Fig. \ref{fig:WarpXvsEDIPIC}) diverge due to the violation of the condition $\Delta t <0.2 \omega_{pe}^{-1} \simeq 2\times 10^{-11} \; s$. The latter condition is a standard requirement for the explicit PIC numerical stability \cite{BirdsallBook}. To achieve the agreement between EDIPIC and WarpX for $R=100$, as shown in Figs. \ref{fig:WarpXvsEDIPIC}, the time and space resolution in WarpX simulations were increased by a factor of four. The number of macroparticles was kept the same for a faster simulation. At the same time, the parameters for EDIPIC simulations remained the same for all values of $R$.

\begin{figure}[htp]
\centering
\captionsetup{justification=raggedright,singlelinecheck=false}
\includegraphics[width=1\textwidth]{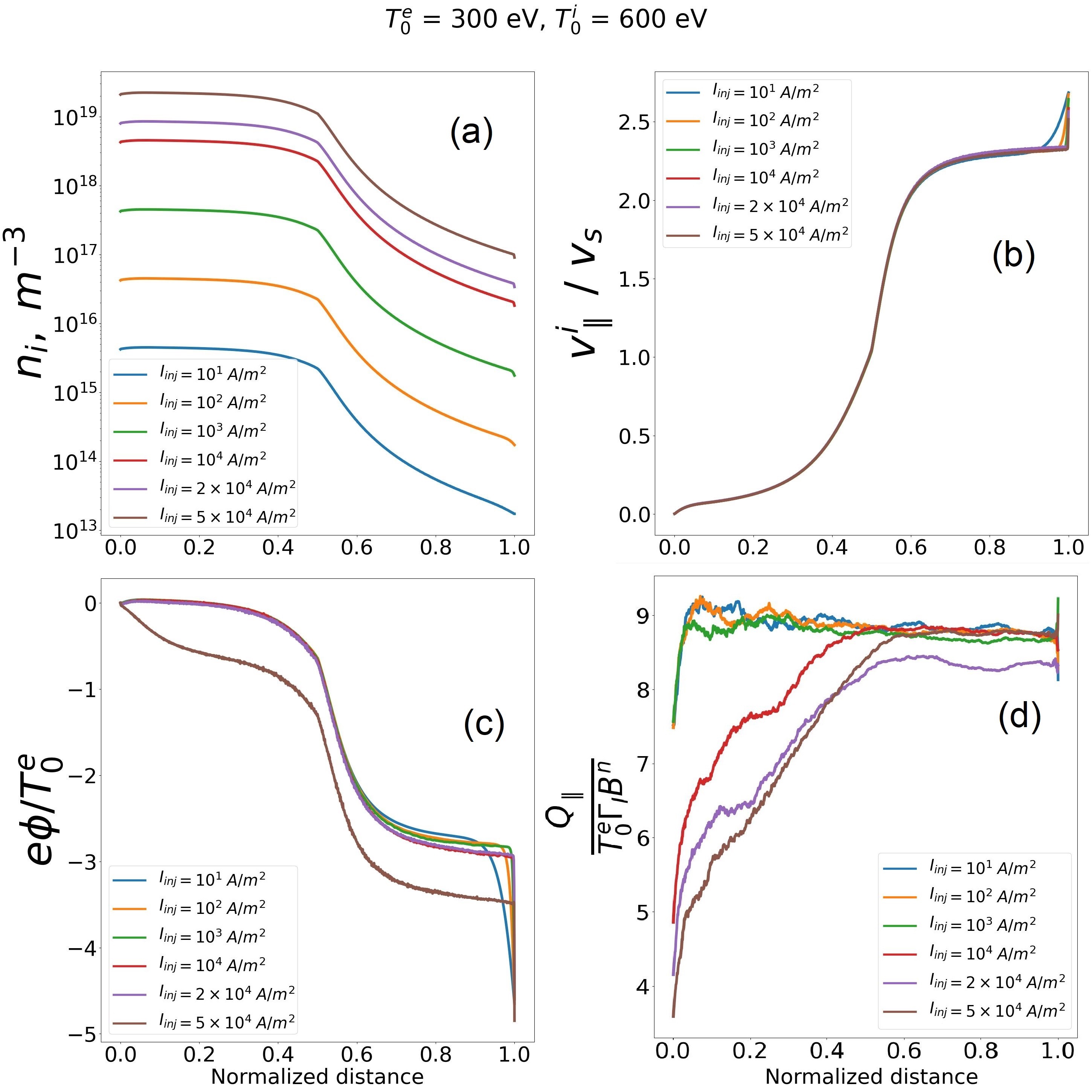}
\caption{Implicit simulations with EDIPIC for different injection currents for the $R=10$, $K=50$ mirror were performed with $\Delta t = 2.5 \times 10^{-11} \; s^{-1}$ and $\Delta z = 10^{-3} \; m$, which were kept fixed for all cases. Comparison of ion concentration (a), ion velocity (b), potential (c), and energy flux (d). The potential profile in (c) for the $I_{inj} = 5\times 10^4 \; A/m^2$ shows large temporal oscillations that do not average out over the time frame accepted for all other cases; (d) the energy conservation starts to deteriorate for the current density above $I_{inj} = 10^3 \; A/m^2$ corresponding to $\Delta z/\lambda_{De}\simeq 5$ and $\omega_{pe} \Delta t  \simeq 0.9$. Note that global density, potential, and ion velocity retain their "universal" profiles even for much larger time steps and grid size.}
\label{fig:Implicit}
\end{figure}

To further test the implicit algorithm of EDIPIC, we simulate different injection rates with the same spatial and time resolution as in base case $\Delta t = 2.5 \times 10^{-11} \; s^{-1}$ and $\Delta z = 10^{-3} \; m$, Figs. \ref{fig:Implicit}. At the low injection flux of $I_{inj}= 10$ $A/m^{2}$, the maximum density in the mirror (before the throat) is $n \simeq 4.2\times 10^{15}$ $m^{-3}$, thus giving $\Delta z/\lambda_{De}\simeq 1.9$ and $\omega_{pe} \Delta t \simeq 0.09$. The plasma density scales linearly with the injection flux, so at the largest injection flux we tested, $I_{inj}=5\times 10^4$ $A/m^{2}$, the plasma density reaches $n\simeq 2.1\times 10^{19}$ $m^{-3}$, Fig. \ref{fig:Implicit}a. At this density, our simulation is performed with $\Delta z/\lambda_{De}\simeq 37$ and $ \omega_{pe} \Delta t \simeq 6.5$.

\begin{figure}[htp]
\centering
\captionsetup{justification=raggedright,singlelinecheck=false}
\includegraphics[width=1\textwidth]{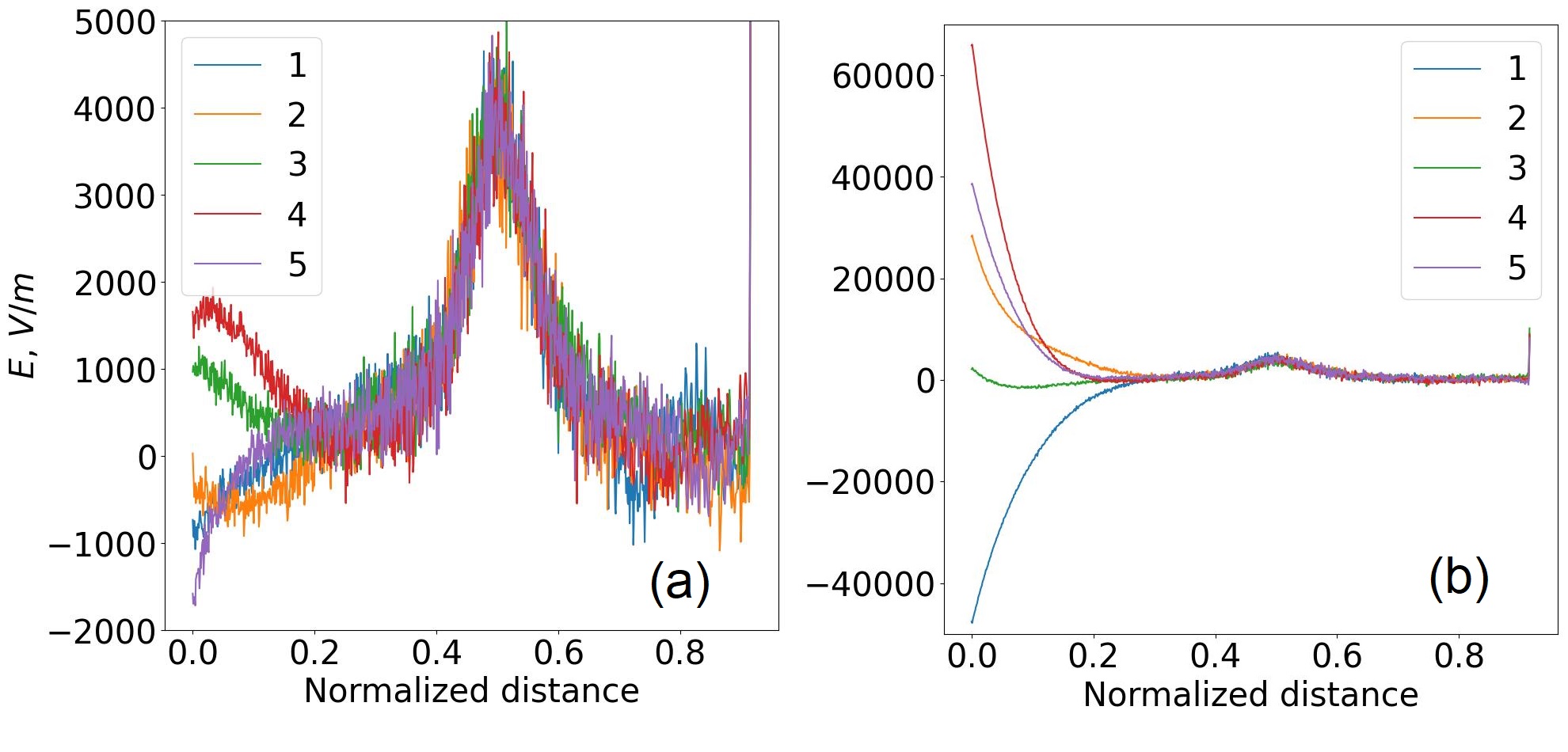}
\caption{Oscillations of the electric field in large density simulations with implicit EDIPIC. (a) Five consecutive snapshots of the electric field for $I_{inj} = 2\times 10^4 \; A/m^2$. Each snapshot is averaged with a moving window of the width 100$\Delta z$ to reduce the spatial noise. Note the large amplitude oscillations observed in the converging part of the mirror near the plasma source; (b) Five consecutive snapshots of the electric field for $I_{inj} = 5\times 10^4 \; A/m^2$. Note that the width of the regions with large amplitude oscillations is increased compared to case (a).}
\label{fig:Waves}
\end{figure}

One can see from Figs. \ref{fig:Implicit}a and b, that with the increase in the injection flux, the density amplitude rescales while global profiles of plasma density, potential, and ion velocity remain unchanged. The sheath size at the right boundary (visible in the potential and ion velocity profiles) decreases, corresponding to the rescaled densities. However, as it is seen in Fig. \ref{fig:Implicit}c, at the largest tested injection rate of $I_{inj} = 5 \times 10^4 \; A/m^2$, the irregular fluctuations in the potential profile appear. They are not smoothed out by our standard averaging procedure done over 300 snapshots uniformly spaced over 9 microseconds (the averaging window for the implicit case was increased for this set of simulations). Large fluctuations at $I_{inj} = 5 \times 10^4 \; A/m^2$ are apparent (under accepted averaging parameters) in the potential profile but are not visible in the ion density nor in the ion velocity. Less apparent oscillations exist at the lower values of the current as well. The oscillations occur in the region of largest density, as Fig.  \ref{fig:Waves}a shows oscillations of the electric field for $I_{inj}=2 \times 10^4 \; A/m^2$ and Fig.  \ref{fig:Waves} for $I_{inj}=5 \times 10^4 \; A/m^2$.  Figs.  \ref{fig:Waves} show five consecutive snapshots of the electric field. The snapshots are averaged with a moving window of the width $100\Delta z$ to reduce noise. Figs.  \ref{fig:Waves}a and b show that the amplitude of the oscillations (between the snapshots) and the width of the region where the oscillations are localized increases with the increase of plasma density.  Fig.  \ref{fig:Implicit}d shows that the energy conservation starts to deteriorate for the currents larger than $I_{inj}=10^3 \; A/m^2$, corresponding to the density larger than $n\simeq 4.2\times 10^{17}$ $m^{-3}$.

Therefore, the EDIPIC implicit scheme allows us to overstep the Debye length and the electron plasma frequency conditions roughly by a factor of 10 but retain good energy conservation. These results are in agreement with general recommendations for the direct implicit scheme (with $D_1$ spatial smoothing \cite{CohenJCP1989}) that conserve energy as long as $\omega_{pe} \Delta t \simeq 30 $ for $\sqrt{T_e / m_e} \Delta t \simeq 0.3 \Delta z$ \cite{CohenJCP1989}. In our case, we have $\sqrt{T_e / m_e} \Delta t \simeq 0.18 \Delta z$. It is interesting to note that even for considerably larger time steps and spatial grid sizes the global density, velocity, and potential profiles do not change and retain the sheath physics while the energy conservation may deteriorate, as seen in Fig.  \ref{fig:Implicit}d. 
Additionally, the behavior of the numerical heating/cooling is non-monotonical with changes in time and spatial steps. As one can see in Fig.  \ref{fig:Implicit}d, the energy conservation is the best for an intermediate density corresponding to the current $I_{inj}=10^3 \; A/m^2$, the green line in Fig.  \ref{fig:Implicit}d. This behavior is reminiscent of that described in Ref. \onlinecite{CohenJCP1989}. Simulations of this scale typically require 2-3 days on the Beluga (The Digital Research Alliance of Canada) cluster with 160 CPUs.

\section{Effects of collisions on trapped electrons and resulting  modifications of the potential profiles and electron distribution function}

The important role of collisions on energy losses in open mirror systems has long been recognized and studied theoretically, through simulations, and experiments. \cite{Pastukhov_1974,KonkashbaevJETP1978,RyutovFST2005,CohenJCP1989,SkovorodinPPR2019,SkovorodinAIP2016,SoldatkinaAIP2016}. A recent work \cite{RosenArxiv2024} has considered the roles of intricate details of the collision operator on plasma confinement in mirror geometry. The comprehensive study of such effects requires a detailed analysis of the role of collisions both in the plasma source and the expander region, which is beyond the scope of our paper. Here, we address the role of collisions and the related effects of trapped electrons in the expander region on the plasma potential and electron pressure anisotropy. For our base case, with $T^i_0 = 600 \; eV$, we include electron-neutral (e-n) elastic collisions with a uniform background of Hydrogen atoms across the entire length of the system. The cross-section for the process was taken from Ref. \onlinecite{MorganDatabase}. The collisions are modeled using the Monte Carlo null-collision method as perfectly elastic collisions. In simulations, the collision frequency is calculated based on the energy dependence of the cross-section, thus becoming a function of the electron energy varying along the nozzle. To characterize different collisional regimes, we use $0.7T^e_0$  for the total electron temperature as an effective value which is equal to the average value in the fully collisional regime in the expander. The same $0.7T^e_0$ value is used to estimate an average bounce frequency, giving $\nu_b=v^e_{th}/L  = 7.2 \times 10^6 \; s^{-1}$ for $L=1.2$ m. These parameters are summarized in \cref{table:coll_table}. 

\begin{figure}[htp]
\centering
\captionsetup{justification=raggedright,singlelinecheck=false}
\includegraphics[width=1\textwidth]{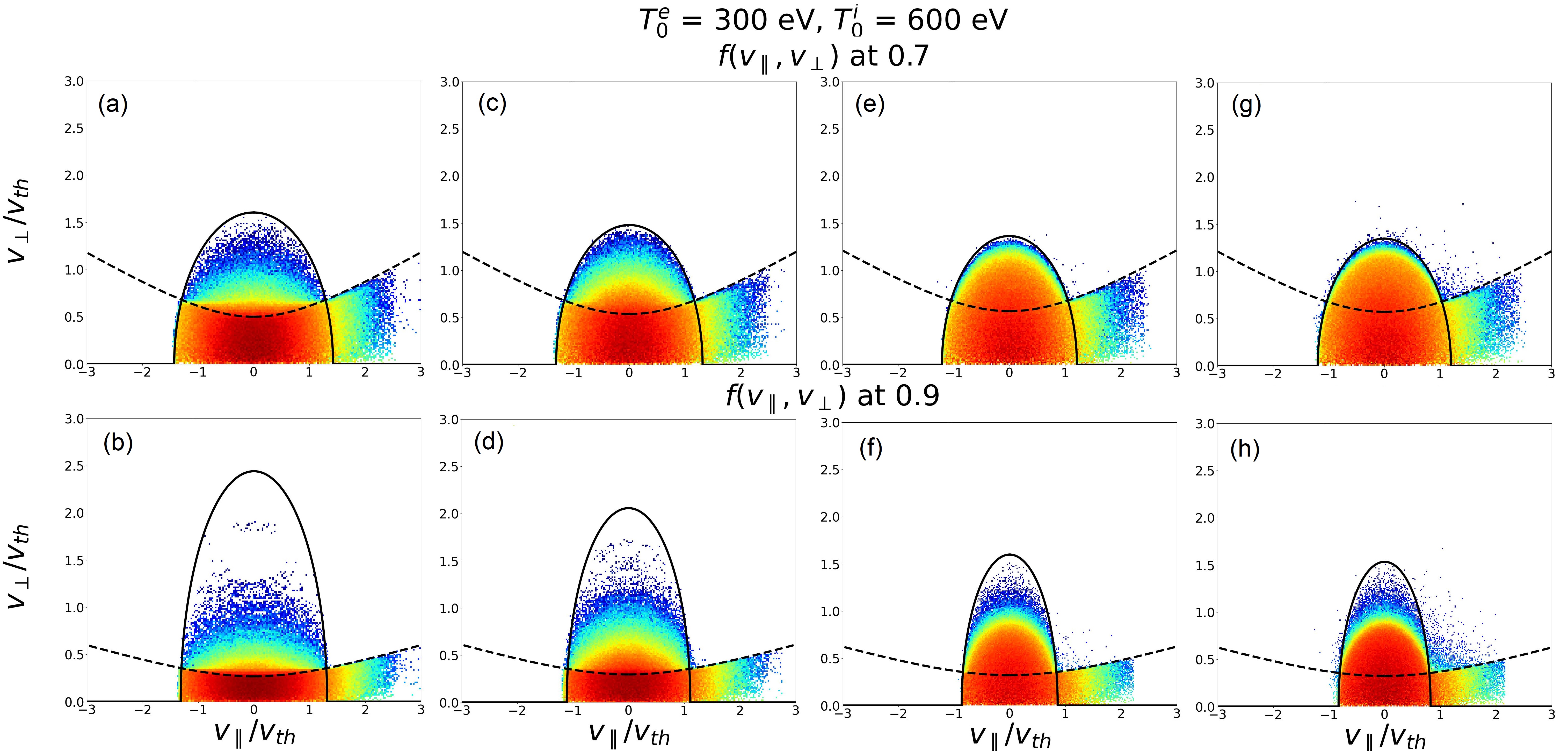}
\caption{Electron distribution function for two locations  in the expander $z/L=0.7$ (top) and $z/L=0.9$ (bottom). Corresponding ratios of the number of trapped particles  to the total: for  $\nu_{en}/\nu_b=1.3\times10^{-3}$  are $22\%$ and $29\%$, respectively in (a) and (b), 
 $\nu_{en}/\nu_b=1.3\times10^{-2}$ --  $46\%$ and $62\%$, in  (c) and (d); $\nu_{en}/\nu_b=1.3\times10^{-1}$, -- $51\%$ and $70\%$ in (e) and (f);  $\nu_{en}/\nu_b=1.3$ --  $53\%$ and $72\%$, in (g) and (h). }
\label{fig:distr_coll}
\end{figure}

The main effect of collisions is the isotropization of the electron distribution function and the related trapping of electrons in the expander.  In Fig. \ref{fig:distr_coll}, electron distribution functions are shown for two locations and different collision frequencies. As expected, the number of trapped particles increases with the collision frequency, and the trapped region shrinks as a result of the concomitant modification of the potential profile.

\begin{figure}[htp]
\centering
\captionsetup{justification=raggedright,singlelinecheck=false}
\includegraphics[width=1\textwidth]{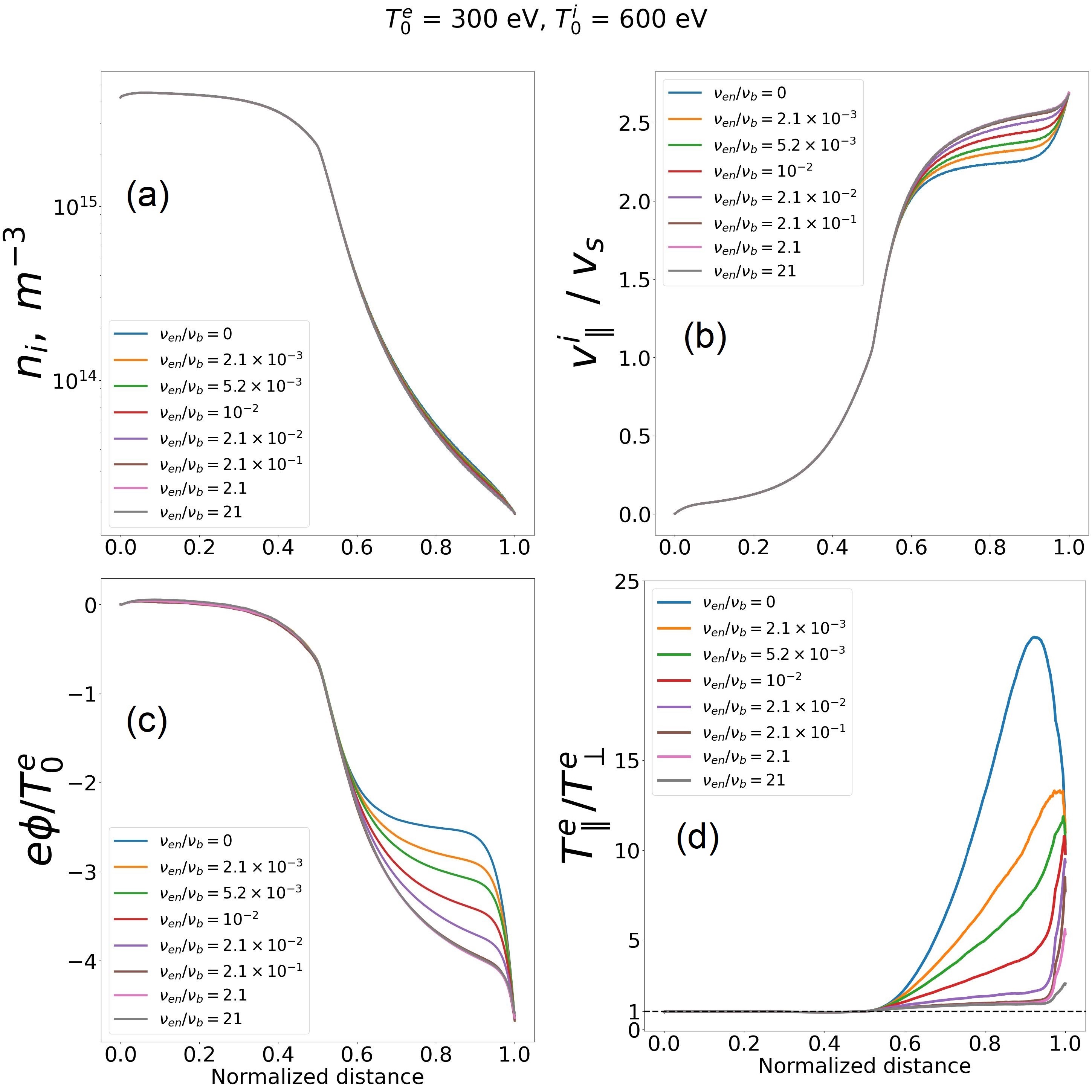}
\caption{Effects of  e-n collisions on plasma density (a), ion velocity (b), and potential profiles (c). The sub-figure (d) shows the  anisotropy of the electron temperature --the ratio of the parallel to perpendicular temperature -- for different collisions frequencies. The base case parameters with $T^i_0 = 600 eV$ and $R=10,K=50$ are used.}
\label{fig:conc_coll}
\end{figure}

\begin{figure}[htp]
\centering
\captionsetup{justification=raggedright,singlelinecheck=false}
\includegraphics[width=1\textwidth]{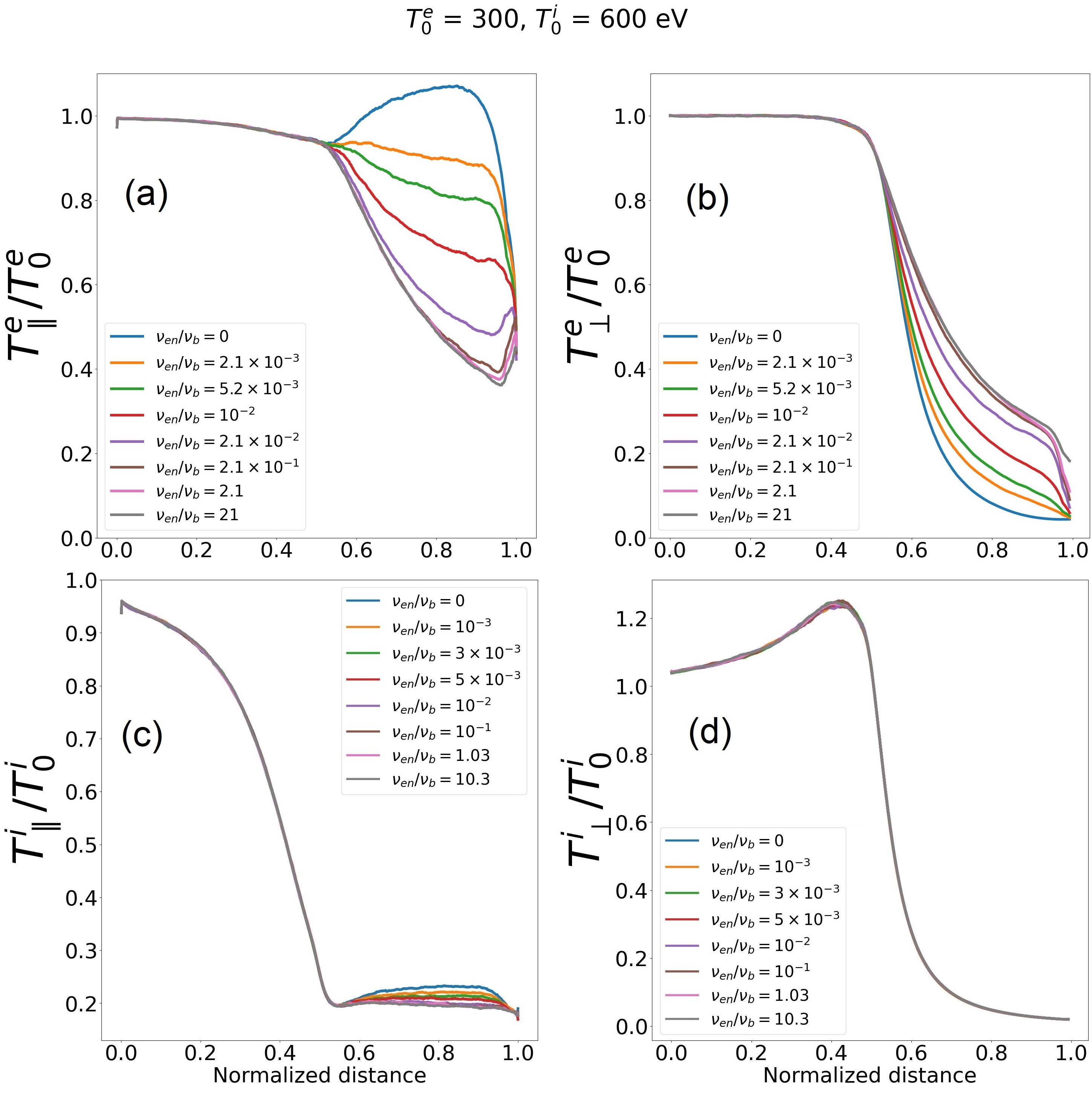}
\caption{Effects of  e-n collisions on the electron, (a) and  (b); and ion, (c) and( d), temperature profiles. }
\label{fig:temp_coll}
\end{figure}

The profiles of the parallel and perpendicular electron and ion temperatures for different collisionalities are shown in Fig. \ref{fig:temp_coll}.   The results in Figure \ref{fig:conc_coll} show that collisions, despite being present throughout the whole system, do not affect the region before the throat of the magnetic nozzle; they are apparent only in the expansion part. The collisions modify the shape of the ion velocity (b) and potential profile (c), but not the total potential drop or the plasma density profile (a). The plasma potential becomes more Boltzmann-like (see Fig. \ref{fig:Kin_boltz_conc}). However, collisions nearly equalize the parallel and perpendicular temperatures for highly collisional regimes, $\nu_{en}/\nu_b$ $>0.1$, which continuously decrease with the expansion and the decrease of the magnetic field,  see Figs. \ref{fig:temp_coll} a and b. For large collisionalities, the potential variations shift away from the absorbing plate toward the maximum of the magnetic field (without changes in the total potential drop across the nozzle) with simultaneous isotropization and a decrease in the temperature.   These results are generally in agreement with the results of  Ref. \onlinecite{ZhouPSST2021}.

\begin{figure}[htp]
\centering
\captionsetup{justification=raggedright,singlelinecheck=false}
\includegraphics[width=1\textwidth]{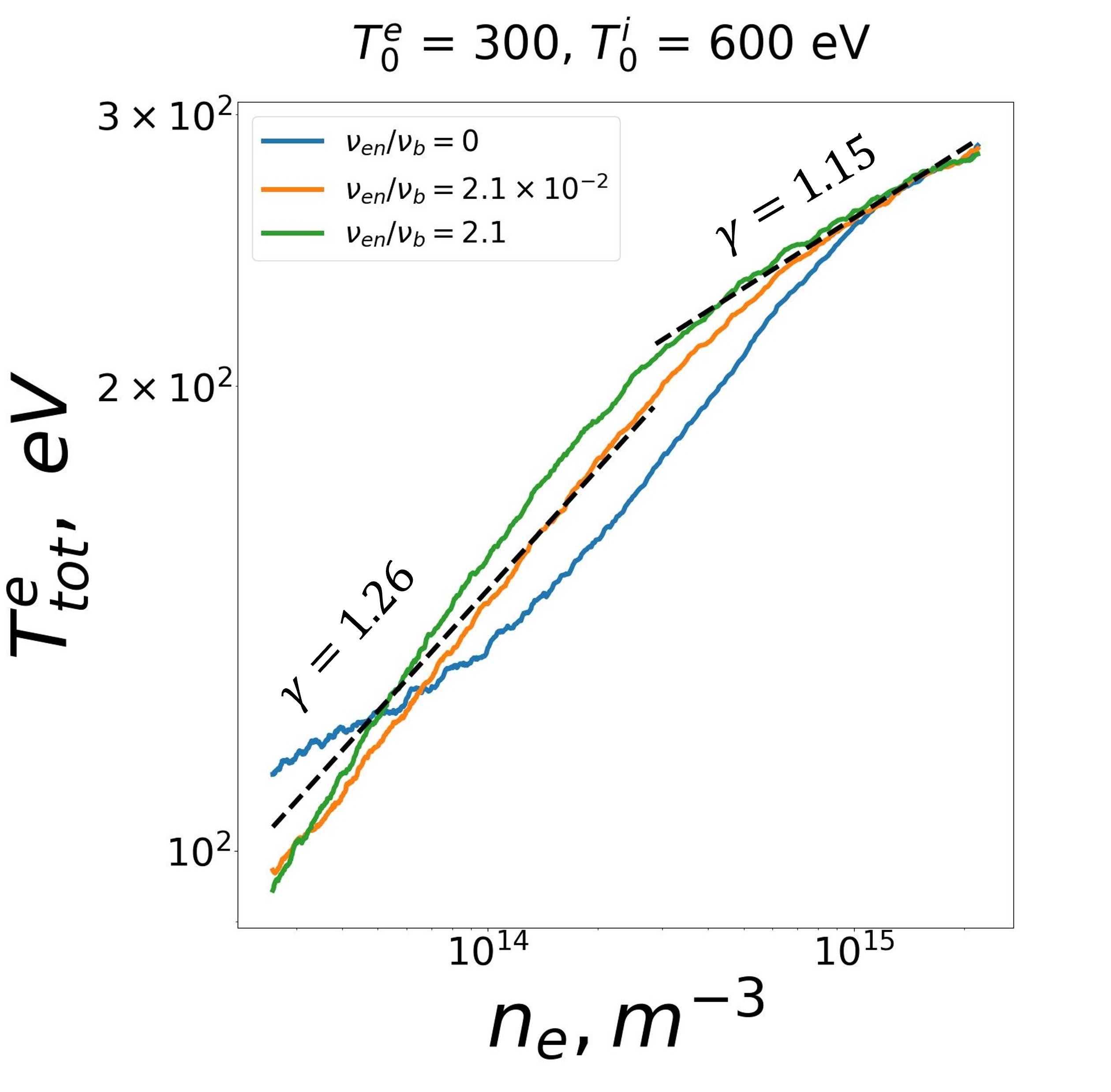}
\caption{Polytropic index $\gamma$ for different collisionalities calculated  from the total electron temperature ($T^e_{tot} = (T^e_\parallel + 2T^e_\perp)/3$) as a function of the electron density,  $n_e$,  in  the  expansion region from $z/L=0.5$ to $z/L=0.9$, excluding sheath region. Note that the value of $\gamma$ changes from about 1.17 near the maximum of the magnetic field, to the larger value of about 1.26 (or slightly larger for large collisionality) at the exit of the nozzle.}   
\label{fig:gamma}
\end{figure}

The electron cooling is a result of plasma expansion, the related conversion of thermal energy into the kinetic energy of ions (via the electric field), and electron heat conductivity. Often, electron cooling is characterized by the polytropic constant $\gamma$, for the isotropic pressure $p\simeq n^\gamma$.  Such characterization is theoretically limited. Its validity depends on the details of the electron distribution function, e.g. degree of pressure isotropization, i.e. isotropic  $p=(2p_\perp+ p_\Vert)/3 $ versus anisotropic $p_\perp \neq p_\Vert$ pressure, the magnitude of heat fluxes and other possible factors, e.g., electric and magnetic fields \cite{LePoP2010,EgedalPoP2013}. The polytropic relation is one of the simplest assumptions of the equation of state for a general moments closure problem in weakly collisional regimes. Nevertheless, it is often used to characterize electron cooling in experiments and modeling \cite{LittlePRL2016,ZhangPRL2016, TakahashiPRL2018,TakahashiPRL2020,CorreyeroPSST2019,ZhouPSST2021}.  In our simulations,  the polytropic index  varies along the nozzle as is shown in Fig. \ref{fig:gamma}, and is not too sensitive to the collisionality (varies between $\gamma=1.2$ for the collisionless case and $\gamma=1.26$ for the highly collisional case near the nozzle exit). Experiments under various conditions report values from almost isothermal $\gamma \simeq 1$ to adiabatic $\gamma=5/3$ and even larger, as discussed and summarized in Ref. \onlinecite{KimJunePSST2023}.   We would like to emphasize that the notion of the polytropic index is limited in scope; for example, the effects of pressure anisotropy are outside the polytropic equation of state.


\begin{nolinenumbers}
\begin{table}[h]
\centering
\caption{Physical parameters for electron-neutral elastic collisions for Hydrogen (H) gas}
\begin{tabular}{|l||*{8}{c|}}\hline
\rowcolor{Gray}
\backslashbox[7cm]{Quantity name}{H concentration, $10^{20}\;m^{-3}$}
&\makebox[1.5em]{$0$}
&\makebox[1.5em]{$0.01$}
&\makebox[1.5em]{$0.025$}
&\makebox[1.5em]{$0.05$}
&\makebox[1.5em]{$0.1$}
&\makebox[1.5em]{$1$}
&\makebox[1.5em]{$10$}
&\makebox[1.5em]{$100$}\\\hline\hline
Neutral pressure, mTorr      & $0$  & $0.03$ & $0.08$ & $0.16$ & $0.31$ & $3.1$ & $31$ & $310$  \\\hline
Averaged collision frequency, MHz, $\nu_{en}$ & $0$ & $0.02$ & $0.05$ & $0.09$ & $0.19$ & $1.9$ & $19$ & $190$ \\\hline 
Collision-to-bounce frequency ratio, $\nu_{en}/\nu_b$  & $0$  & $0.001$ & $0.003$ & $0.005$ & $0.01$ & $0.103$ & $1.03$ & $10.3$ \\\hline 
 \end{tabular}
\label{table:coll_table}
\end{table}
\end{nolinenumbers}

\section{Summary and discussion}
We have presented the effective drift-kinetic PIC model to study plasma flow, acceleration, and energy transport in the geometry of the magnetic mirror. Despite of its limitations related to paraxial approximation, this model provides useful insights into the physics of plasma acceleration and axial energy losses in the magnetic mirror and magnetic nozzle.

 We have employed implicit EDIPIC code to study the regimes approaching realistic plasma parameters with high density,  realistic lengths, and electron/ion mass ratio.   
The results from EDIPIC and WarpX codes well agree in the explicit regime with small time steps and small grid size. The EDIPIC implicit algorithm allows the simulations to overstep the Debye length and electron plasma frequency conditions. We find that the energy is well conserved up to $\Delta z/\lambda_{De} \simeq 10$ and $\omega_{pe} \Delta t = 6.5$. The simulations with even larger time and spatial size remain stable and show the same "universal" profiles of plasma density, velocity, and potential, thus opening the possibilities for practical simulations of realistic dimensions and plasma densities. We have found that practical limits of the implicit algorithm exist roughly at $\Delta z/\lambda_{De}\simeq 37$ and $ \omega_{pe} \Delta t \simeq 6.5$ when the profile of averaged potential starts to depart from the "universal" profile.

We have shown directly in our kinetic model how the finite ion temperature in the plasma source increases plasma acceleration out of the mirror and axial energy losses. Full kinetic results, in particular, the heat flux, are compared with the results obtained in the hybrid formulation with kinetic ions, isothermal Boltzmann electrons, and quasineutrality. Furthermore, the ion kinetic results are compared with the results of the fluid closure model where the collisionless heat fluxes are calculated from the extended (higher order, Grad type) hydrodynamic equations for the time evolution of the heat fluxes. It is shown that the kinetic results reasonably agree with the results from the extended hydrodynamic closure model. It is suggested that such a fluid closure model can also be useful to characterize the thermodynamic properties of the electron component.
 
We have studied the relative contributions of the heat fluxes to the energy transport and energy conversion in each component. Also, the formation and development of the anisotropies in the electron and ion distribution function, and related electron and ion pressure anisotropies, are demonstrated at different locations along the mirror. The heat fluxes for the parallel and perpendicular energy (both for electron and ion components), responsible for the development of the pressure anisotropies, are explicitly determined here. These results show the role of the heat fluxes at different locations and can be used to describe thermodynamic properties \cite{TakahashiPRL2020, KimPSST2018NJP2018, KimJunePSST2023} of the compressing and expanding flow of plasma in the mirror, i.e., the transitions between limiting cases of the isothermal and two-pressure adiabatic regimes that may exist at different regions in the mirror. We find that the heat flux magnitudes are largest in the region of the maximum magnetic field, Figs. \ref{fig:Kinc_Boltz_heat}. 

 The energy balance is investigated in detail, revealing the energy conversion between electrons and ions, ultimately defining the total axial energy losses in the mirror. The global integral relations are derived that relate the energy input/loss in each species to the total potential drop along the mirror, Eqs. (\ref{pi}) and (\ref{pe}). It is shown that the total potential drop across the mirror (including the sheath) saturates at large expansion and remains largely constant for large $K$. This behavior is different from the logarithmic dependence of the total potential drop in the hybrid model with isothermal electrons.

We have previously argued that quasineutral plasma flow and acceleration \cite{SmolyakovPoP2021, JimenezPoP2022, SaboPoP2022} in the converging-diverging magnetic field configurations, such as the mirror and magnetic nozzle, are robustly constrained by the regularization condition at the sonic point, i.e., at the point where the plasma flow velocity is equal to the local ion sound velocity. The regularization condition defines a unique solution with the velocity that follows the magnetic field profile as shown analytically \cite{SmolyakovPoP2021} and numerically with a hybrid model using Boltzmann isothermal electrons and drift-kinetic ions \cite{JimenezPoP2022}. Here, the results of the full kinetic model for ions and electrons, including the Poisson equation, demonstrate similar robustness of the global profiles of plasma density, ion velocity, and potential profiles in the converging part of the mirror. In particular, we find that in this region   there exists a strong coupling of the potential with the magnetic field profile similar to what is obtained in simple analytical theory and suggested by the experiments \cite{SunPRL2005, LongmierPSST2011}. 
However, in the diverging part of the mirror, the full kinetic results diverge from those predicted by the isothermal model. We note that isothermal Boltzmann distribution for electrons is an asymptotic limit of zero electron flux, i.e. the result of full reflection of all electrons supporting infinite effective electron heat conductivity subsequently leading to the infinite potential drop which requires an  infinite energy source upstream. In practice,  the ion acceleration by the electric field occurs at the expense of the electron thermal energy therefore leading to the electron cooling \cite{MartinezSPoP2015}. This is the main reason for the deviation of the kinetic results from the hybrid model with isothermal Boltzmann electrons in the expander region. This region remains quasineutral but the electron pressure becomes anisotropic due to a strong decrease in the perpendicular temperature. Thus,  the electron cooling is anisotropic: the perpendicular pressure drops much faster which occurs near the maximum of the magnetic field.  The effect of the electron cooling is pronounced further in the non-quasineutral downstream region near the absorbing plate where ions gain additional acceleration.  

We find that in our simulations with $T_0^e=300$ eV electron temperature from the source, the axial energy losses per one ion Eq. (\ref{ec2}) range from $\eta\simeq 6$ (for low ion energy, $T_0^i \ll T_0^e$) to $\eta \simeq 9$ for higher ion energy of $T_0^i=600$ eV. For a moderate expansion ratio of $K=50$, roughly one $T_0^e$ is carried by electrons overcoming the sheath barrier, and the rest is by the accelerated ions that get their energy from the electrons via the electric field.  These values are close to the energy losses reported in the mirror experiments \cite{SoldatkinaAIP2016, SoldatkinaNF2020, GotaNF2021} and theoretical publications \cite{GuptaPoP2023}. The energy losses are weakly sensitive to the values of K, but confinement improves with R, e.g. as it appears in the increased density in the source region. 


It is expected that collisions and trapped electrons affect the plasma potential and energy losses, especially when electrons are cold, e.g. as produced in the expander region by ionization and Secondary Electron Emission (SEE) from the absorbing plate\cite{RyutovFST2005, SkovorodinAIP2016, SkovorodinPPR2019}. 
In our collisionless simulations, we observe about 20\% of trapped particles due to transient processes and (perhaps) PIC numerical noise. This number is similar to the 25\% result observed in other collisionless simulations, e.g.  Ref. \onlinecite{ZhouPSST2021}.

Collisions scatter particles into the trapped region. To study such effects, we performed a separate series of simulations including electron-neutral collisions with a uniform background of neutral gas, allowing electrons from the source to become trapped due to scattering.  This study shows that the main effects of collisions are the isotropization of electron pressure and related modifications of the potential profile, which becomes more Boltzmann-like, but with a continuous decrease in electron temperature in the expander. The electric field in the expander becomes larger, but the total potential drop remains the same, causing the sheath width to decrease. This behavior is similar to that observed in Ref. \onlinecite{ZhouPSST2021}.    

Full isotropization of electron pressure is observed only for extremely large values of the collision frequency, $\nu_{en}/\nu_b \geq 1$. For realistic parameters of interest, the electron-electron collision frequency is much smaller compared to the bounce frequency.  Thus, the average e-e Coulomb collision frequency $\nu_c$ for $T_e = 300 \; eV$ and $n = 4\times 10^{14} \; m^{-3}$  $\nu_c \approx 3 \; s^{-1}$, the bounce frequency $\nu_b=v_{th}/L \approx  10^7 \; s$, and dimensionless collisionality parameter is ${\hat \nu}=\nu/\nu_b \approx 3 \times 10^{-7}$.   Therefore, our results, as well as those of  Ref. \onlinecite{ZhouPSST2021},  indicate that in the absence of sources of cold electrons in the expander region, particle trapping directly from the source only marginally modifies plasma parameter profiles. We also note that for small values of the dimensionless collisionality parameter in the range of ${\hat \nu}=\nu/\nu_b \leq  10^{-3}$ one still expects significant anisotropy of the electron pressure, as shown in Fig. \ref{fig:conc_coll}d.

Our study neglects the effects of particle collisions on the distribution function formed  in the source\cite{Pastukhov_1974}. Particle collisions critically affect the overall energy losses by modifying the distribution function injected into the mirror, at $z'=0$.  In our model, it was assumed that particles in the confinement volume acquire Maxwellian distribution because the confinement time is large compared to the collision time.  From the perspective of energy flux, the full Maxwellian source, as it is assumed in our work, is the worst case and therefore represents an upper limit on energy losses through the mirror from the confinement volume.  For weakly collisional plasmas of interest, as per the estimates above, the assumption of the Maxwellian source is not valid for particles in the loss cone region -- these particles are lost immediately as they are created and heated resulting in the depletion of the high energy tail of the distribution function (both for electrons and ions). Such depletion and associated effects on the energy losses were studied previously and remain an area of active research for the open mirror systems\cite{Pastukhov_1974,KonkashbaevJETP1978, SoldatkinaNF2020,SoldatkinaPoP2017,SoldatkinaAIP2016,SkovorodinAIP2016,SkovorodinPPR2019} but rarely discussed for propulsion applications. 

Thus, the effects of collisions go  beyond the problem of trapped particles alone and require separate studies including the collisional mechanisms in plasma source, and additional sources of cold electrons in the expander such as ionization and SEE. These topics are outside of the scope of the present work and left to a separate study.

\acknowledgments{This work is supported in part by NSERC Canada.  The computational resources were provided by the Digital Research Alliance of Canada. This research used the open-source particle-in-cell code WarpX https://github.com/ECP-WarpX/WarpX, primarily funded by the US DOE Exascale Computing Project. Primary WarpX contributors are with LBNL, LLNL, CEA-LIDYL, SLAC, DESY, CERN, and TAE Technologies. We acknowledge all WarpX contributors.}


\section*{Data availability}
The data that support the findings of this study are available from the corresponding author upon reasonable request.

\clearpage 
\appendix

\section{Poisson equation in the paraxial model}

In the paraxial approximation, two-dimensional effects are effectively included in the one-dimensional model by averaging over the variable cross-section of the narrow flux tube near the symmetry axis. Consider the two-dimensional Poisson equation:
\begin{equation}
\nabla \cdot \boldsymbol{E=}4\pi e(n_{i}-n_{e}).
\end{equation}
Averaging the left-hand side of this equation in the poloidal plane for small $r$, one obtains:
\begin{eqnarray}
\left\langle \nabla \cdot \boldsymbol{E}\right\rangle  &\equiv &\frac{1}{\pi
\delta ^{2}}\int_{0}^{\delta }2\pi rdr\left( \frac{\partial }{\partial z}%
E_{z}+\frac{1}{r}\frac{\partial }{\partial r}rE_{r}\right) \simeq \frac{%
\partial }{\partial z}E_{z}+\frac{2}{\delta ^{2}}\int_{0}^{\delta }dr\frac{%
\partial }{\partial r}rE_{r}  \notag \\
&=&\frac{\partial }{\partial z}E_{z}+\frac{2}{\delta ^{2}}\left.
rE_{r}\right\vert _{0}^{\delta }.
\label{avd} 
\end{eqnarray}
Assuming $\boldsymbol{E=}E_{\Vert }\boldsymbol{b}$ and using the expression for the radial magnetic field in the paraxial approximation, one obtains
\begin{equation}
E_{r}=E_{\Vert }B_{r}/B=-E_{\Vert }\frac{r}{2}\frac{1}{B}\frac{\partial B_{z}%
}{\partial z}.
\end{equation}%
This gives from (\ref{avd}) 
\begin{equation}
\left\langle \nabla \cdot \boldsymbol{E}\right\rangle =\frac{\partial }{%
\partial z}E_{z}-E_{z}\ \frac{1}{B}\frac{\partial B_{z}}{\partial z},
\label{}
\end{equation}%
which is identical to Eq. (\ref{parp}).

\section{Boundary conditions}

Our simulations aim to represent half of the symmetric system with a plasma source in the center. Therefore, having $E=0$ and $\phi=0$ at the left wall is a natural choice of physically realistic boundary conditions for one-dimensional simulations, leaving the potential at the left wall free and expecting that the potential drop along the mirror axial direction is established self-consistently from the quasineutrality and ambipolarity (zero current) conditions. In practical finite-length applications, such as the magnetic mirror, the sheath will naturally occur at the collector wall. Therefore, another option for boundary conditions (BC) is $\phi=0$ at the left wall and the floating wall condition at the right wall. The floating wall at $z=L$ results in a finite value of the electric field establishing zero net current. The total current at the left wall is zero due to the equal rates of the injection of electrons and ions and the reflux of all particles. The floating wall forms the sheath near the right wall (the collector sheath), thus regulating the electron current to establish global quasineutrality, control the energy flux, and ensure stationary current-free plasma acceleration.

\begin{figure}[htp]
\centering
\captionsetup{justification=raggedright,singlelinecheck=false}
\includegraphics[width=1\textwidth]{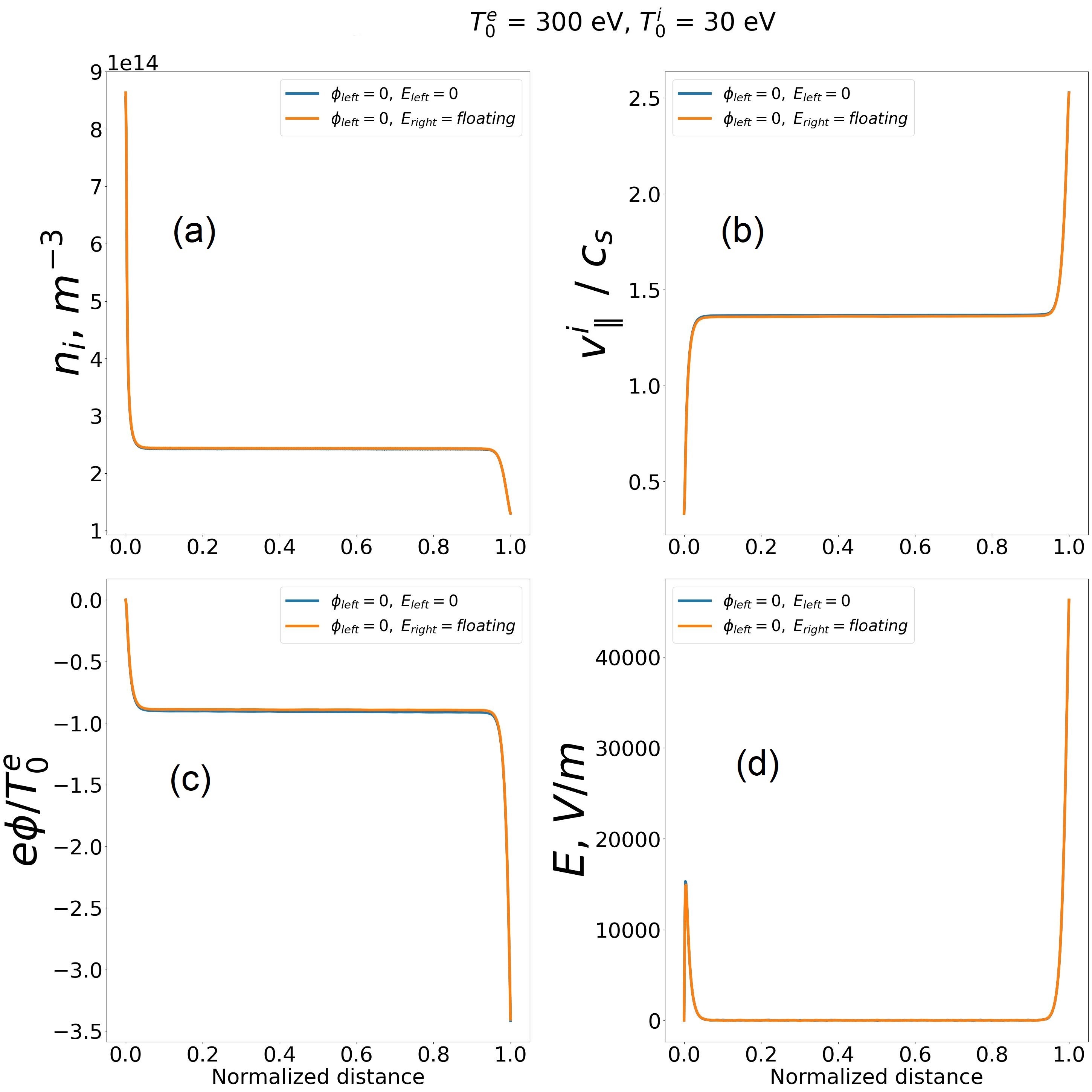}
\caption{Time-averaged plasma parameters for simulations with two different boundary conditions and a uniform magnetic field for ion concentration (a), ion velocity (b), potential (c), and electric field (d). Note the formation of the source sheath similar to Ref. \onlinecite{SchwagerPFB1990}.}
\label{fig:ShwagBound}
\end{figure}

We have performed simulations employing both types of boundary conditions for the case of a uniform magnetic field as in Ref. \onlinecite{SchwagerPFB1990}. The comparison in Figure \ref{fig:ShwagBound} shows that stationary solutions obtained by using the floating wall BC and the BC with zero potential and electric field, $\phi=0$, $E=0$ at the left wall, $z=0$, are nearly identical.

\begin{figure}[htp]
\centering
\captionsetup{justification=raggedright,singlelinecheck=false}
\includegraphics[width=1\textwidth]{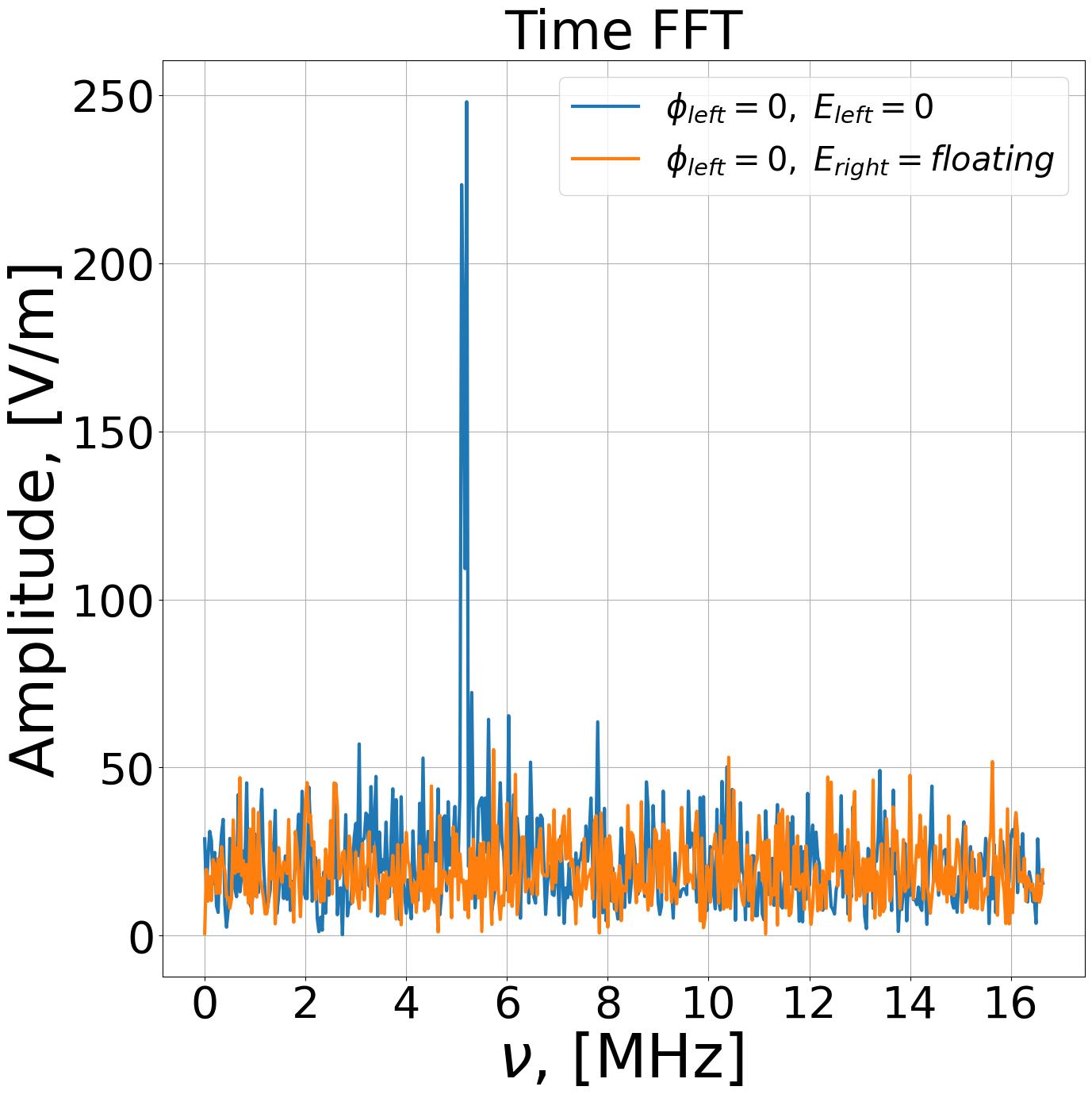}
\caption{The Fast-Fourier transform (FFT) for the electric field at $z=L/2$ in simulations with two different boundary conditions.}
\label{fig:Fourier}
\end{figure}

One has to note that the solution with the floating wall BC is less noisy and requires a shorter time window for averaging. In the simulations with the zero potential and electric field BC at $z=0$, the noise appears as a strong peak at the frequency above or around the ion plasma frequency, as shown in Fig.  \ref{fig:Fourier} for the Fast-Fourier transform (FFT) of the electric field at $z=L/2$. It appears that these oscillations are the result of numerical instability typical for open flow boundary conditions \cite{Brieda2006, Li_2019, JambunathanIEEE2020}.

We have also tested both boundary conditions for the case of the mirror converging-diverging magnetic field. Again, the simulations with the zero potential and electric field BC at $z=0$  show large noise levels. It is remarkable that despite large oscillation the averaged plasma parameters profiles for both BC remain very similar. However, a much longer time averaging window is required for the mirror magnetic field compared to the case of the uniform field.  


In a half-infinite plasma expanding in a vacuum, such as in propulsion applications, quasineutrality and ambipolarity set up the potential that asymptotically becomes constant at a sufficiently large distance (formally at infinity) from the source, such that no sheath occurs. In finite-length simulations purporting to describe semi-infinite plasma, some sort of logical sheath and current control are typically used to prevent the formation of the sheath at the collector side \cite{Brieda2006, Li_2019, JambunathanIEEE2020, ChaconJCP2024}.

\section{Extended magnetohydrodynamic equation for ions taking into account the collisionless heat fluxes}

Extended fluid equations for ions take into account the pressure anisotropy, heat fluxes in the energy balance, and the time-dependent evolution equations for the heat fluxes. These equations can be obtained by truncating general moment equations of the Grad hierarchy \cite{OraevskiPP1968}, or directly as the heat moments of the drift kinetic equation (\ref{dke}), and calculating the fourth-order moments with a two-temperature Maxwellian distribution. Such equations were earlier derived in Ref. \onlinecite{OraevskiPP1968}, see also Ref. \onlinecite{Khazanov_book}, and also in Refs. \onlinecite{SnyderPoP1997, FundamenskiPPCF2005}. Neglecting all collisional effects, such equations can be written in the form:

 \begin{equation}
    \frac{\partial n}{\partial t} + V_{\parallel}\frac{\partial n}{\partial z} + n\frac{\partial V_{\parallel}}{\partial z} - nV_{\parallel}\frac{\partial \ln B}{\partial z} = 0,
\label{continuity-eq}
\end{equation}
\begin{equation}
    m_{i}n\left(\frac{\partial V_{\parallel}}{\partial t} + V_{\parallel}\frac{\partial V_{\parallel}}{\partial z}\right) = -en\frac{\partial \phi}{\partial z} - \frac{\partial p_{i\parallel}}{\partial z} + \left(p_{i\parallel} - p_{i\perp}\right)\frac{\partial \ln B}{\partial z},
\label{momentum-eq}
\end{equation}
\begin{equation}
\begin{split}
    \frac{\partial p_{i\parallel}}{\partial t} + V_{\parallel}\frac{\partial p_{i\parallel}}{\partial z} + 3p_{i\parallel}\frac{\partial V_{\parallel}}{\partial z} + \frac{\partial q_{i\parallel}}{\partial z} & + \left(2q_{i\perp} - q_{i\parallel} - V_{\parallel}p_{i\parallel}\right)\frac{\partial \ln B}{\partial z} = 0,
\label{parallel-pressure-eq}
\end{split}
\end{equation}
\begin{equation}
\begin{split}
    \frac{\partial p_{i\perp}}{\partial t} +V_{\parallel}\frac{\partial p_{i\perp}}{\partial z}  + \frac{\partial q_{i\perp}}{\partial z} + p_{i\perp}\frac{\partial V_{\parallel}}{\partial z} - 2\left(q_{i\perp} + V_{\parallel}p_{i\perp}\right)\frac{\partial \ln B}{\partial z} = 0,
\end{split}
\label{perpendicular-pressure-eq}
\end{equation}
\begin{equation}
\begin{split}
    \frac{\partial q_{i\parallel}}{\partial t} + V_{\parallel}\frac{\partial q_{i\parallel}}{\partial z} + 4q_{i\parallel}\frac{\partial V_{\parallel}}{\partial z} - V_{\parallel}q_{i\parallel}\frac{\partial \ln B}{\partial z} + 3 \frac{p_{i\parallel}}{nm_{i}}\frac{\partial p_{i\parallel}}{\partial z} - 3\frac{p_{i\parallel}^2}{n^2m_{i}}\frac{\partial n}{\partial z}  = 0,
\end{split}
\label{parallel-heat-flux-eq}
\end{equation}
\begin{equation}
\begin{split}
    \frac{\partial q_{i\perp}}{\partial t} & + V_{\parallel}\frac{\partial q_{i\perp}}{\partial z} + 2q_{i\perp}\left(\frac{\partial V_{\parallel}}{\partial z} - V_{\parallel}\frac{\partial \ln B}{\partial z}\right) + \frac{p_{i\parallel}}{nm_{i}}\frac{\partial p_{i\perp}}{\partial z} - \frac{p_{i\parallel}p_{i\perp} }{n^2m_{i}}\frac{\partial n}{\partial z} + \frac{p_{i\perp}}{nm_{i}}\left(p_{i\perp} - p_{i\parallel}\right)\frac{\partial \ln B}{\partial z} = 0.
\end{split}
\label{perpendicular-heat-flux-eq}
\end{equation}

\bibliographystyle{unsrt}
\bibliography{ref}

\clearpage
\newpage



\end{document}